\documentclass[3p]{elsarticle}

\usepackage{preamble}
\usepackage{gensymb}
\usepackage{natbib}
\usepackage{float} 
\usepackage{makecell}
\usepackage{array}
\usepackage{mathtools}
\usepackage{color,soul}
\usepackage[none]{hyphenat}

\title{In Silico Evaluation of Cardiac Tissue-Engineered Patch Interventions}

\begin{document}

\newcolumntype{C}[1]{>{\centering\arraybackslash}m{#1}}


\author[um]{John Patrick Sayut, Jr. ~\texorpdfstring{\corref{CA}}}
\ead{jsayut@umich.edu}

\author[um]{Javiera Jilberto}
\ead{jilberto@umich.edu}

\author[um]{Mia Bonini}
\ead{mbonini@umich.edu}

\author[kcl]{Marc Hirschvogel, Ph.D.}
\ead{marc.hirschvogel@ambit.net}

\author[um]{Will Zhang, Ph.D.}
\ead{willwz@umich.edu}

\author[um,kcl]{David A. Nordsletten, Ph.D.}
\ead{nordslet@umich.edu}

\address[um]{
    Department of Biomedical Engineering,
    University of Michigan, 
    1221 Beal Avenue, Ann Arbor, MI, USA, 48109-2102}
\address[kcl]{
    Division of Biomedical Engineering and Imaging Sciences,
    Department of Biomedical Engineering,
    King's College London,
    4th Floor, Lambeth Wing, St. Thomas' Hospital,
    Westminster Bridge Road, London, UK SE1 7EH}

\cortext[CA]{Corresponding author: John Patrick Sayut, Jr., North Campus Research Center, Building 400, 1600 Huron Parkway, Ann Arbor, MI 48109. e-mail: jsayut@umich.edu phone: +1 (321) 506-4589}

\begin{abstract}
    Myocardial infarction significantly degrades heart function, and current treatments can bring forth serious cost and complications including blood clots and infections. To improve the current state of treatment, researchers are developing tissue patches from induced-pluripotent stem cells that can be incorporated into the heart, improving organ function after a myocardial infarction. These tissue patches include surface patches, attached to the epicardium of the heart, and thick transmural patches that replace the infarcted region. However, little is known about the impact of cardiac tissue patches on pump function in a patient's heart. In addition, it is not clear what patch structural properties - such as active stress generation, muscle fiber alignment, or material stiffness - may best augment existing heart tissue. Computational modeling can be used to examine different implementations and patch properties, illuminating the mechanical impact of cardiac tissue patches in the beating heart. In this work, we computationally implement different cardiac tissue patches to understand benefits of particular patch types and properties. We find that in transmural cardiac tissue patches, both activation and fiber alignment improve function. A transmural patch generating 10\% of healthy active stress can increase stroke volume by 18\%, and higher generated active stress in a circumferential muscle fiber orientation can recover stroke volume by over 50\%. Furthermore, we find that surface cardiac tissue patches can enhance heart function slightly despite limiting diastolic filling, especially when fibrotic thinning has occurred. These conclusions identify broad design goals for the engineering of cardiac tissue patches to improve heart function after a myocardial infarction.

\end{abstract}

\bigskip

\begin{keyword}
    cardiac biomechanics;
    heart modeling;
    cardiac fibrosis; 
    regenerative medicine;
    computational simulation;
    computational modeling;
\end{keyword}

\maketitle

\section{Highlights}
\begin{itemize}
    \item Biventricular heart models allow us to study cardiac tissue patches
    \item Higher generated active stress in cardiac tissue patches improves function
    \item Native fiber alignment in cardiac tissue patches also improves function
    \item Higher active stress in patches increases functional benefit of native fibers
    \item Transmural cardiac tissue patches outperform surface patches
\end{itemize}

\section{Introduction}

%

\par Myocardial infarctions (MI) impact 9.5\% of individuals globally above age 60, creating a region of heart tissue that is stiff, non-contractile, and incapable of healing \cite{tunstall-pedoe_myocardial_1994, hinderer_cardiac_2019, salari_global_2023}. 
MI often leads to deteriorating heart function and eventually chronic heart failure, requiring treatment either through a heart transplant or a left ventricular assist device, both of which carry complications including blood clots and infections \cite{cameli_donor_2022, estep_risk_2015, llerena-velastegui_efficacy_2024}. Because of this, tissue engineers are searching for stem cell interventions that could improve the functionality of the patient's native heart tissue rather than replicating this tissue  \cite{tenreiro_next_2021}. 
Researchers hypothesize that injecting stem cells directly into fibrotic tissue might promote regeneration, but this approach increases the risk of arrhythmogenic events \cite{liu_human_2018, marchiano_gene_2023, nakamura_pharmacologic_2021}.
An alternative approach is to engineer stem-cell-derived cardiac tissue patches, surgically attached to a patient to improve ventricular contraction \cite{tsan_physiologic_2021, shadrin_cardiopatch_2017, jebran_engineered_2025}. 
Through surgery, these innovative patches could be added to the surface of the heart or incorporated transmurally to replace scarred myocardium \cite{zimmermann_multilayer_2023}. They could be constructed to have different cardiac muscle fiber alignments, mechanical behaviors, and contractile properties, determined by how the selected protocol impacts maturity and tissue structure \cite{tenreiro_next_2021}. These different patch properties represent engineering tradeoffs that are critical to effective patch design.
Understanding how different patch design parameters impact a patient's heart function is essential for tissue engineering researchers to construct the best possible patch.

\par To develop cardiac tissue patches, researchers innovate through experimental methods. Many tissue engineering teams study cardiac contraction after implantation, building tissue patches and characterizing contraction in animal models \cite{jabbour_vivo_2021, guan_transplantation_2020, lou_cardiac_2023, gao_large_2018, jebran_engineered_2025}. However, the contractile properties of many cardiac tissue patches reflect neonatal behavior, limiting the ability of researchers to explore the functional benefit of more mature active stress generation in vivo \cite{tenreiro_next_2021}. Similarly, researchers have made great strides in constructing cardiac muscle fibers that can support cardiac tissue constructs \cite{jabbour_vivo_2021, guan_transplantation_2020, lou_cardiac_2023, gao_large_2018, jebran_engineered_2025}. However, manipulating the muscle fibers of cardiac tissue patches to match patient-specific native fiber orientation is a significant design challenge. Experiments to quantify patch property engineering tradeoffs are both expensive to run and difficult to compare between tissue engineering teams, representing a tissue engineering challenge that is well served by computational investigation.

\vspace{5mm}

\par Computational research studying stem-cell-derived cardiac tissue patches can reduce the difficulty of patch experiments and allow for the direct comparison of patch properties.
For modeling heart function in silico generally, multiphysics solvers incorporate physics-based mathematical relationships, allowing researchers to capture phenomena such as fluid-structure interaction and circulatory responses \cite{syed_modeling_2023, hirschvogel_monolithic_2017}.
Researchers have applied computational modeling to develop virtual experimental cohorts, to investigate the value of potential interventions, and even to build virtual clinical trials \cite{miller_implementation_2021, hirschvogel_silico_2019, ohara_personalized_2022, niederer_creation_2020}. 
Cardiac fibrosis after MI has been biomechanically assessed through finite element modeling, illustrating how fibrotic material becomes stiffer and more isotropic as it remodels  \cite{janssens_post-infarct_2023, mojsejenko_estimating_2015}. 
Investigators have also used computational models to study the impact of cardiac tissue patches, investigating how cardiac muscle fiber alignment of a surface patch alters mechanical and functional outcomes in an idealized left ventricle \cite{janssens_impact_2024, janssens_role_2025}. In this project, we seek to use a biventricular model to investigate how activation, fiber alignment, and mechanical stiffness impact both epicardial and transmural cardiac tissue patches.
Computational modeling empowers the rapid functional assessment of cardiac tissue patches, allowing us to meaningfully detail the cardiac tissue patch design space for design teams.

\vspace{5mm}

\par The objective of this paper is to characterize engineering tradeoffs key to improving cardiac function through surface and transmural tissue-engineered patches. 
By applying a representative baseline computational model that considers the passive solid biomechanics of the myocardium, we will solve for myocardial active stress throughout the cardiac cycle and the acute response of the cardiovascular system.
Cardiac tissue patch attachment method and patch structural properties will be varied, as well as the left ventricular morphology in the case of the surface patch. 
This approach allows us to compare the impact of these variables on the heart's pump function.
We find that both activation and fiber alignment in transmural cardiac tissue patches improve pump function and that surface cardiac tissue patches can slightly enhance heart function despite limiting diastolic filling, especially when fibrotic thinning has occurred post-MI. 

\vspace{5mm}

\section{Methods}

\par We seek to model cardiac tissue-engineered patches to understand the impact of patch properties on heart function. This requires a solid mechanical model for the heart that reflects the microstructure (Section \ref{s_biomechanical}), as well as material behavior (Section \ref{s_representative}). We will illustrate the impact of post-MI fibrosis by simulating a fibrotic region, and then we will simulate both the application of transmural and surface patches, developing geometries for each model (Sections \ref{s_bv_geometry} and \ref{s_patch_geometry}). To monitor how these patches influence the remaining cardiovascular system, we incorporate an acute systemic response by coupling a 0D lumped parameter representation of the cardiovascular system to the biventricular geometry (Section \ref{s_multiphysics}). We determine the different patch structural properties that must be incorporated to meaningfully sweep the parameter space, and how to incorporate them (Section \ref{s_variations_patch}). We then identify functional metrics that can be incorporated to illustrate the impact of simulation results on patient health, and finally execute the simulations (Sections \ref{s_functional_metrics} and \ref{s_simulations}). 
\subsection{Biomechanical Modeling} \label{s_biomechanical}
\par To model the solid mechanics of the heart, we use a continuum mechanics formulation \cite{holzapfel_nonlinear_2010}. We desire a constitutive material model for the myocardium that captures biomechanical behavior and also reflects what is known of the microstructure. To develop a relationship between stress and deformation, we define reference domain coordinates $\mathbf{X}$ and displacement $\mathbf{u}(\mathbf{X},t)$.  In the expressions for these material models, the deformation gradient $\mathbf{F}=\nabla_0\mathbf{u}+\mathbf{I}$ indicates how points in domain $\mathbf{X}$ deform on the heart mesh, and the volumetric deformation of an element is written as $J=\text{det}\mathbf{F}$. The right Cauchy-Green tensor is defined as $\mathbf{C}=\mathbf{F^T}\mathbf{F}$ which represents how vectors on the heart mesh deform. The stress tensor is comprised of four components \cite{nordsletten_viscoelastic_2021}. The second Piola-Kirchoff stress tensor is written as
\begin{equation}
    \mathbf{S}=\mathbf{S_v}+\mathbf{S_e}+\mathbf{S_p}+\mathbf{S_a}
\end{equation}
\par  The model $\mathbf{S_v}$ reflects the passive hyperelastic response of the myofibers, $\mathbf{S_e}$ reflects the Neo-Hookean hyperelastic response of the ground matrix, $\mathbf{S_p}$ reflects the hydrostatic pressure response of the tissue, and $\mathbf{S_a}$ reflects cardiac muscle fiber active stress that develops in the material \cite{nordsletten_viscoelastic_2021}. 
\par The relationship for the microstructural material response $S_v$ is written in Equation \ref{viscoe} \footnote{In this case, we apply the fractional viscoelastic material model from \cite{nordsletten_viscoelastic_2021} but strictly consider the hyperelastic response, where $\alpha=0$}. 
\begin{equation}
\label{viscoe}
\begin{multlined}
    \mathbf{S_v}= b_{ff}(W_1 I_{ff}-1) \mathbf{\hat{e}_f} \otimes \mathbf{\hat{e}_f} + b_{ss}(W_1 I_{ss}-1) \mathbf{\hat{e}_s} \otimes \mathbf{\hat{e}_s} + b_{nn}(W_1 I_{nn}-1) \mathbf{\hat{e}_n} \otimes \mathbf{\hat{e}_n} 
    \\ + W_2 ( b_{fs} I_{fs} \operatorname{sym}(\mathbf{\hat{e}_f} \otimes \mathbf{\hat{e}_s}) + b_{fn} I_{fn} \operatorname{sym}(\mathbf{\hat{e}_f} \otimes \mathbf{\hat{e}_n}) + b_{sn} I_{sn} \operatorname{sym}(\mathbf{\hat{e}_s} \otimes \mathbf{\hat{e}_n}))
\end{multlined}
\end{equation}
\begin{equation}
    W_1=e^{b_1 (I_C - 3)} \quad W_2=e^{b_2 (I_{fs}^2 + I_{fn}^2 + I_{sn}^2)}
\end{equation}
\begin{equation}
    I_{ab}=\mathbf{C}:\operatorname{sym}(\mathbf{\hat{e}_a} \otimes \mathbf{\hat{e}_b}), \quad  a,b \in {f,s,n}
\end{equation}
\par The vector $\mathbf{\hat{e}_a}$ is a unit vector in the reference microstructural direction defined by $a \in \{f,s,n\}$. In this case, the symmetric transformation is defined as $\operatorname{sym}(\mathbf{A})=\frac{1}{2} (\mathbf{A} + \mathbf{A^T})$. Material parameters are indicated by $b_1$, $b_2$, and $b_{ij}  \quad  i,j \in {f,s,n}$.
\par The Neo-Hookean hyperelastic response $\mathbf{S_e}$ of the ground matrix is written in Equation \ref{neo-h} where $a$ represents the Neo-Hookean material parameter.  
\begin{equation}
\label{neo-h}
    \mathbf{S_{e}}=\frac{a}{J^{2/3}}(\mathbf{I}-\frac{\mathbf{C:I}} {3}\mathbf{C}^{-1})    
\end{equation}
The heart's volumetric response is modeled as compressible with the form indicated in Equation \ref{compr}, with compressible parameters $K_1$ and $K_2$. 
\begin{equation}
\label{compr}
    \mathbf{S_p}=\mathbf{C^{-1}}(K_1(J-1)J + K_2\operatorname{log}(J))   
\end{equation}
 Lastly, we apply active stress $\mathbf{S_a}$ using a length-dependent formulation \cite{kerckhoffs_homogeneity_2003}.  The uniform ventricular contraction parameter is identified as $\alpha$ at each time point of the cardiac cycle. This uniform ventricular contraction parameter is separated between the left and right ventricle, as further detailed in Section \ref{s_representative}.
\begin{equation}
\label{eq_active}
   \mathbf{S_a}= \alpha \space \operatorname{tanh}[3.75 (\lambda_f  - \lambda_{f_{min}}) ]^2(\mathbf{\hat{e}_{f}} \otimes \mathbf{\hat{e}_{f}} + \frac{1}{3}\mathbf{I}) \quad \lambda_f > \lambda_{f_{min}}
\end{equation}
\begin{equation}
\label{fiber_active}
    \lambda_f = \sqrt{\mathbf{C :} (\mathbf{\hat{e}_{f}} \otimes \mathbf{\hat{e}_{f}})}
\end{equation}
As can be seen from Equation \ref{eq_active} and \ref{fiber_active}, active stress is applied primarily along the fiber direction, alongside secondary isometric contraction representing known fiber dispersion \cite{tangney_novel_2013, kassab_microstructurally_2016}. Minimum stretch in contraction is selected as $\lambda_{f_{min}} = 0.8$.
\par To incorporate microstructure-based constitutive models of the heart, we must develop a muscle fiber field that matches the heart geometry of our representative dataset. This field is developed via a rule-based method that defines Laplace problems across the heart tissue and solves for appropriate fiber angles \cite{bayer_novel_2012} \cite{doste_rule-based_2019}. For our mesh, we define fiber angles with respect to the circumferential that vary from $-60\degree$ to $60\degree$ in the left ventricle and $-25\degree$ to $90\degree$ in the right ventricle, from the epicardium to the endocardium \cite{doste_rule-based_2019}. This fiber field is illustrated in Figure \ref{fig:geometry}E and further described in Section \ref{s_bv_geometry}.

\par The transient conservation of mass and momentum for the solid material is governed by Equation \ref{strongform}. 

\begin{equation}
\label{strongform}
    \rho \frac{\partial^2 u}{\partial t^2} = \nabla_\mathbf{X} \cdot (\mathbf{FS})
\end{equation}

\par In this case, $\frac{\partial \cdot}{\partial t}$ is the Lagrangian derivative taken with respect to the reference domain coordinates $\mathbf{X}$. 

\subsection{Cardiac Material Law Parameterization} \label{s_representative}

\par We seek a mechanical response representing a patient suffering from a myocardial infarction to understand the difference between healthy function, diseased function, and interventions. The Klotz curve identifies a relationship between end-diastolic pressure and volume across species and disease \cite{klotz_single-beat_2006}. To parameterize our solid mechanics model, we identify material parameters to match patient data using a Klotz curve optimization protocol. This protocol is further described in the Appendix.

\par Table \ref{tab:material_parameters} indicates the material parameters that we use to develop a representative
model for this problem. The first nine parameters specify the material model as described in Section \ref{s_biomechanical}. In the Klotz optimization protocol, we determine independent linear stiffness values for the right and left ventricles (RV and LV). The parameter $R$ is the stiffness multiplier for the RV, operating on all linear stiffness parameters $b_{ab_{RV}}=(R) b_{ab_{LV}} \quad  a,b \in {f,s,n} $ and operating similarly on ground matrix parameter $a$. The parameters defining the stiffness of fibrosis are also shown in the table, based on data found in \cite{mcgarvey_temporal_2015}. To represent this data in our hyperelastic model, we define the scaling $\kappa$ as a linear multiplier for all linear hyperelastic parameters in $S_v$, for example $b_{ab}=\kappa b_{ab_0}$ for $ a,b \in {f,s,n} $. The fiber direction scaling $k_f$ adds to this multiplier exclusively in the fiber direction $b_{ff}=(\kappa+k_F)b_{ff_0}$. We identify separate contraction variables for the left and right that must be found based on the ventricular pressure-volume relationship.

\begin{table}[h]
\centering
\begin{tabular}{|p{1.75cm}|p{2.0cm}|p{10.25cm}|}
\hline
\textbf{Parameter} & \textbf{Value} & \textbf{Description} \\
\hline
$b_1$ & 4.61 [-] & Power law parameter for microstructural hyperelasticity \\ \hline
$b_2$ & 0.533 [-] & Power law parameter for microstructural hyperelasticity \\ \hline
$b_f$ & 0.331 [kPa] & Microstructural parameter, fiber direction \\ \hline
$b_s$ & 0.181 [kPa] & Microstructural parameter, sheet direction \\ \hline
$b_n$ & 0.0827 [kPa] & Microstructural parameter, sheet-normal direction  \\ \hline
$b_{fs}$ & 1.25 [kPa] & Microstructural shear parameter, fiber and sheet directions \\ \hline
$b_{fn}$ & 0.711 [kPa] & Microstructural shear parameter, fiber and sheet-normal directions \\ \hline
$b_{sn}$ & 0.585 [kPa] & Microstructural shear parameter, sheet and sheet-normal directions \\ \hline
$a$ & 0.0841 [kPa] & Neo-Hookean parameter \\ \hline
$R$ & 28.9 [-] & Right ventricle stiffness scaling \\ \hline
$K_1$ & 100 [kPa] & Compressible bulk modulus \\ \hline
$K_2$ & 100 [kPa] & Compressible bulk modulus \\ \hline
$\kappa$ & 10.7 [-] & Fibrosis scaling in all directions \\ \hline
$k_f$ & 1.24 [-] & Fibrosis scaling in the cardiac muscle fiber direction \\ 
\hline
\end{tabular}
\caption{Material parameters of healthy myocardium and fibrotic scaling values}
\label{tab:material_parameters}
\end{table}

\subsection{Biventricular Geometry and Fibrotic Region Development} \label{s_bv_geometry}

\par We aim to develop heart geometries to investigate the cardiac tissue patch parameter space in a clinically representative way. The two necessary geometries will be referred to as the "chronically thinned" left ventricular morphology, and the baseline left ventricular geometry as the "acute fibrotic" left ventricular morphology. The selected representative patient is 63 years old, male, weights 269 $\textbf{lbs}$ and is 76 $\textbf{in}$ tall. As a male at this age with an obese BMI of 32.75, the selected patient possesses several risk factors for a myocardial infarction, making them a representative example \cite{ojha_myocardial_2025}. To develop the acute fibrotic left ventricular morphology, we apply a uNET segmentation algorithm to our 4D computed tomography angiography of the heart \cite{ennis_whole_2021}. We examine and correct the resulting neural network segmentation in 3D Slicer \cite{ennis_whole_2021, fedorov_3d_2012}.
Next, we define a region of fibrosis in the LV wall of the patient's heart geometry. This area has a radius of 15 $\text{mm}$, along with a border zone linearly varying from that 15 mm radius to a 30 $\text{mm}$ radius (Figure \ref{fig:geometry}A). 
With reference to the literature, we determine the relative stiffness between remote healthy tissue and fibrotic tissue \cite{mcgarvey_temporal_2015, mojsejenko_estimating_2015}. In the area that is fully fibrotic, no active stress is produced. In the adjacent border zone, the active stress scaling increases linearly with distance from the fully fibrotic area, with 100\% contraction beyond the 30 $\text{mm}$ radius. We incorporate fibrotic thinning of 37.8\% of total wall thickness in the fibrotic region, reflecting the approximate change in heart left ventricular morphology after one month as described in \cite{mcgarvey_temporal_2015} (see Figure \ref{fig:geometry}B).

\subsection{Patch Attachment Method Geometry Development} \label{s_patch_geometry}

\par
\begin{figure}[H]
\centering
\includegraphics[width=\textwidth]{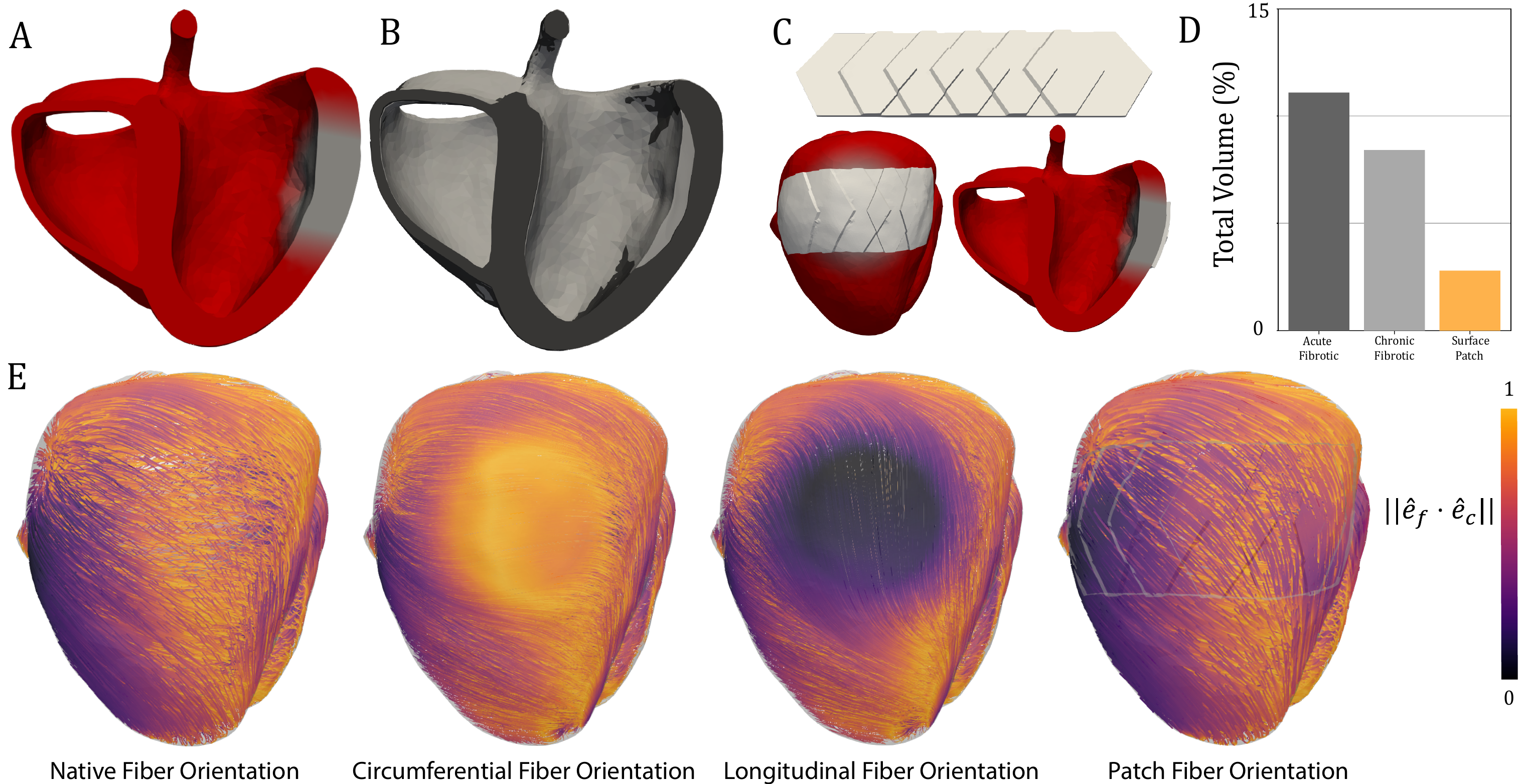}
\caption{Geometry and structure analyzing patch performance. A) The biventricular geometry with healthy myocardium in red and the fibrotic region and border zone in grey. B) The chronically thinned fibrotic left ventricular morphology shown in black with the acute fibrotic left ventricular morphology overlaid in light gray. C) The analytical engineered surface patch geometry and its adherence onto the biventricular heart D) Volume plot showing the volume of the fibrotic regions and the surface patch divided by the total mesh volume for each geometry. In this case, the fibrotic region includes the border zone volume according to its extent of fibrosis. E) From left to right - cardiac muscle fiber orientation for the native fibers, circumferential muscle fibers in the transmural patch, longitudinal muscle fibers in the transmural patch, and the surface patch. $\mathbf{\hat{e}_C}$ refers to the unit vector in the circumferential direction.  
}
\label{fig:geometry}
\end{figure}

 We construct biventricular heart models with geometries that represent relevant patch attachment methods. Namely, we seek to model transmural and surface patch attachment methods. To simulate the transmural patch attachment method, we replace the fibrotic section with healthy constitutive properties and nonzero activation. The transmural cardiac tissue patch is modeled with the stiffness parameters of the native myocardium, except in models where the stiffness structural property is explicitly altered. When a transmural patch is simulated, the fibrotic tissue region is exactly replaced by the cardiac tissue patch \cite{zimmermann_multilayer_2023}. Additionally, in transmural cardiac tissue patches, we seek to investigate differences in functional impact between patches with a native cardiac muscle fiber orientation (including transmural variation) and patches with all cardiac muscle fibers aligned in the same direction and oriented relative to the ventricle. To represent this, we rotate cardiac muscle fibers in the transmural patch to match the corresponding longitudinal or circumferential directions, as identified in Figure \ref{fig:geometry}E. 

\par In the case of a surface patch attachment method, the patch is designed with a thickness of 2mm at the peak (15.1\% of total wall thickness in the fibrotic region), consisting of overlapping cardiac tissue slices as illustrated in Figure \ref{fig:geometry}C \cite{zimmermann_multilayer_2023} \cite{jebran_engineered_2025}. This design is based on a cardiac tissue patch implementation from the BioVAT-HF clinical trial through the German Center of Cardiovascular Research (DZHK) in Germany \cite{noauthor_dzhk_nodate} \cite{gavenis_safety_2023}. More details on this tissue patch and its design can be found in \cite{jebran_engineered_2025} \cite{tiburcy_defined_2017}. 
This surface patch geometry is warped onto the surface of the biventricular model over the region of fibrosis, and is attached to both the acute fibrotic and the chronically thinned fibrotic biventricular morphologies (Figure \ref{fig:geometry}C). The resulting volumes of both transmural and surface cardiac tissue patches relative to the total heart wall volume are shown in Figure \ref{fig:geometry}D. For the surface patch, cardiac muscle fibers were defined to match the epicardial fiber direction, based on the optimal outcome found in \cite{janssens_impact_2024}. Figure \ref{fig:geometry}E illustrates the different cardiac muscle fiber fields for the study, for both transmural and surface patch attachment methods. The selected patch geometries model characteristics of both transmural and surface cardiac tissue patches.

\subsection{Coupling to the Cardiovascular System} \label{s_multiphysics}

\par
\begin{figure}[H]
\centering
\includegraphics[width=0.5\textwidth]{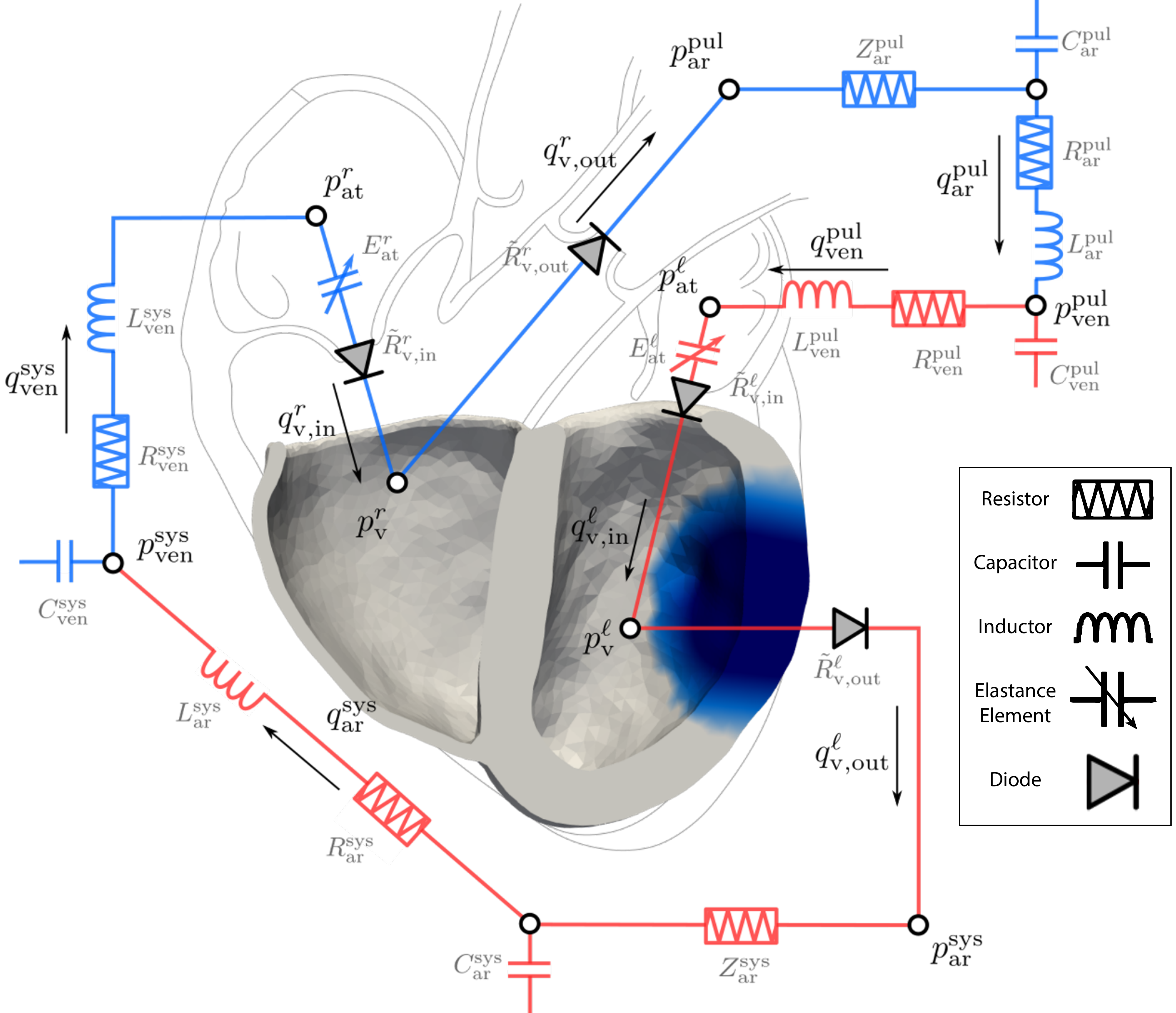}
\caption{3D-0D coupled diagram: Variables represented by the letter $p$ indicate chamber pressure in the 3D ventricles and 0D lumped parameter representation atrial and circulatory elements, letters $R$ indicate resistance of the given vessel to flow for a given pressure (with $Z$ representing a second resistor in the same vessel), letters $E$ indicate a time-varying elastance that allows for blood ejection from arteries, letters $C$ indicate the compliance of a vessel to change flow for a given pressure time derivative, and letters $L$ indicate an inductance element that drives the vessel's resistance to change in flow rate for a given pressure. The circulatory model contains 0D chambers for arterial and venous pressure in both systemic and pulmonary circulatory pathways, with subscript $\text{ven}$ indicating venous circulation, subscript $\text{ar}$ indicating arterial circulation, superscript $\text{sys}$ identifying systemic circulation, and superscript $\text{pul}$ identifying pulmonary circulation. The superscript letters $\text{l}$ and $\text{r}$ indicate ventricular circulation. Adapted from \cite{hirschvogel_monolithic_2017}}
\label{fig:threed_zerod}
\end{figure}

 To illustrate the acute systemic response of the cardiovascular system to changes in the ventricular pressure-volume relationships, we build our computational model from a 0D lumped parameter representation \cite{hirschvogel_monolithic_2017}. The 0D parameters for this study are first defined via echocardiograms, cardiac catheterization, and CT data. The full 0D lumped parameter representation is shown in Figure \ref{fig:threed_zerod}, illustrating 0D circuit elements that represent components of the cardiovascular system. Applying the 0D pressure-volume relationships to the non-fibrotic solid mechanical model for one heart cycle, we determine a unique contraction parameter values for each ventricle over time \cite{miller_implementation_2021}. The geometry and patch structural properties are adjusted for each desired patch model, and we couple this 3D model to the 0D lumped parameter model \cite{hirschvogel_monolithic_2017}. 
The coupled model provides an acute prediction of the cardiovascular system response to patch implantation and associated changes in pressure-flow relationships. 

\subsection{Tissue-engineered Patch Variations} \label{s_variations_patch}

\par
\begin{figure}[H]
\centering
\includegraphics[width=0.5\textwidth]{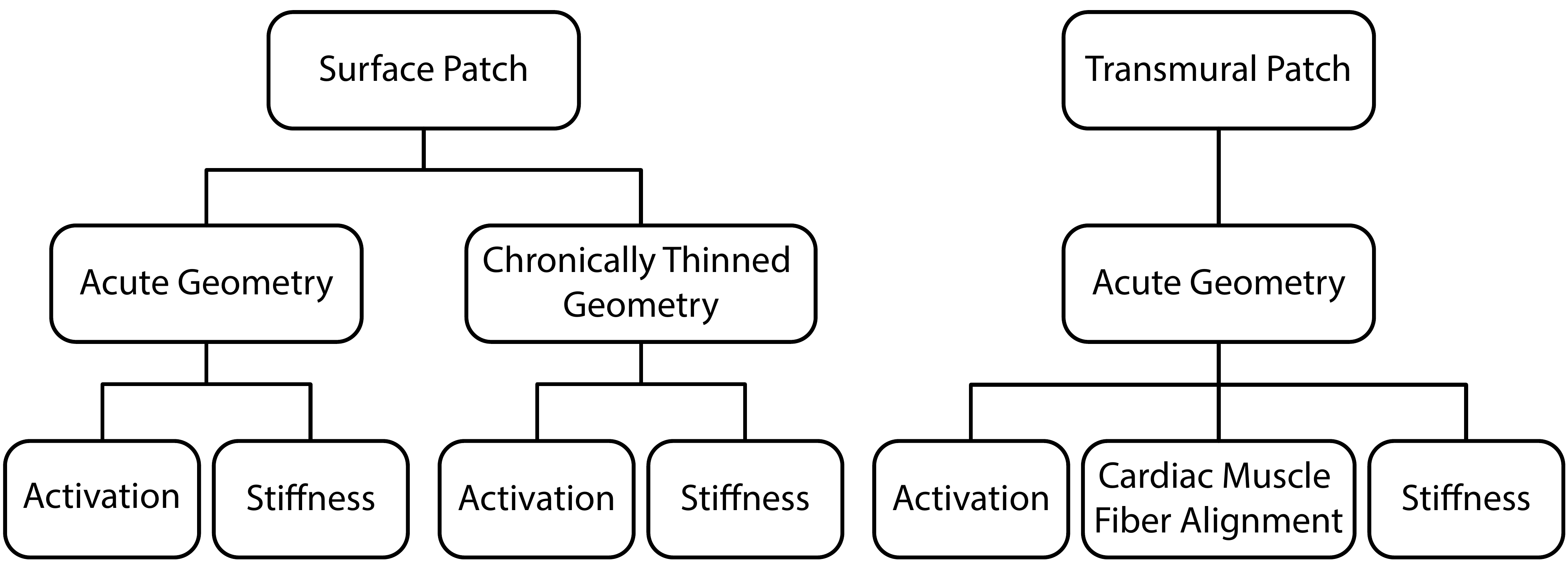}
\caption{Patch properties considered in this study. For applying the patch, we can consider a surface patch that is attached to the epicardium of the LV, and we can also consider a transmural patch that replaces the entire LV wall in place of the fibrotic region. We can consider both "acute" geometries with thick walls and "chronically thinned" geometry in the case of the surface patch. Lastly, for patch structural properties, we consider activation, mechanical stiffness, and cardiac muscle fiber alignment. }
\label{fig:patch_flowchart}
\end{figure}

 For the purpose of examining the patch design parameter space, we identify and implement models showcasing the patch structural properties and patch attachment methods that are critical to intervention design, as shown in Figure \ref{fig:patch_flowchart}. Patch attachment method is of particular concern for engineering teams currently designing cardiac tissue interventions, reflecting different concepts of applying cardiac tissue to patient hearts \cite{jabbour_vivo_2021, gao_large_2018, zimmermann_multilayer_2023, jebran_engineered_2025}. 

\par We identified important patch structural properties from the literature and implement them in our solid mechanical models. These patch structural properties are cardiomyocyte contraction, cardiac muscle fiber orientation, and material stiffness. Contraction is a critical function of mature cardiac tissue, constituting a critical benchmark for cardiac tissue engineering teams \cite{tenreiro_next_2021}. Since it is the primary responsibility of cardiac tissue, tissue engineering teams focus on contraction in evaluating their results \cite{jebran_engineered_2025, jabbour_vivo_2021}. We study cardiomyocyte contraction by a multiplicative scaling on the active stress generated in the patch region during the cardiac cycle. Additionally, the variation of cardiac muscle fiber direction across the heart wall is complex, and requires significant tissue engineering design in the case of transmural patches \cite{zimmermann_multilayer_2023} \cite{bayer_novel_2012}. Cardiac muscle fiber orientation is altered in the transmural patch, varying from the native structure to a circumferential or longitudinal orientation with distinct polarity. Lastly, material stiffness is a significant consideration given the neonatal qualities of some engineered cardiac tissues \cite{tenreiro_next_2021}. Stiffer and softer material properties are generated by scaling $b_{ab}$, with $a,b \in f,s,n$ by values of $1.5$ and $0.5$, respectively, from Equation \ref{viscoe}. Altogether, the patch structural properties represent critical components of cardiac tissue patches that could be used in the clinic in the near future.

\par In addition to patch structural properties, we consider how left ventricular morphology and patch attachment method change the geometry of our system. We examine left ventricular morphology by thinning the endocardial wall of the fibrotic region in the LV. This reflects the chronic fibrotic thinning that is known to occur \cite{mcgarvey_temporal_2015}. We consider the patch attachment method by exploring how transmural and surface patches may perform differently for similar structural properties. Different groups have identified these attachment methods as relevant design choices \cite{zimmermann_multilayer_2023} \cite{lou_cardiac_2023} \cite{jebran_engineered_2025}. Left ventricular morphology and patch attachment method, combined with patch structural properties, together reflect a meaningful sweep of the parameter space for cardiac tissue patch design.

\subsection{Cardiac Function Analysis} \label{s_functional_metrics}

\par In quantifying the impact of different patches on the heart, we seek to understand clincally-based metrics that reflect ventricular function. To do so, we calculate functional metrics employed by clinicians to understand a patient's heart, including stroke volume ($SV$), ejection fraction ($EF$), and stroke work ($SW$) as shown in Equation \ref{functional_metrics}. 

\begin{equation} 
\label{functional_metrics}
    SV = V_{ED} - V_{ES}, \quad EF=\frac{V_{ED}-V_{ES}}{V_{ED}}, \quad SW=\int_{0}^{T} P \frac{dV}{dt} dt
\end{equation}

In this case, $V$ indicates the ventricular volume, with subscripts $ES$ and $ED$ indicating end-systolic and end-diastolic volumes respectively, with $P$ indicating the ventricular pressure trace and $T$ indicating the period of the heart cycle. Additionally, we will examine the peak systolic pressure ($P_{max}^{sys}$) of the heart as an additional indicator of systolic function. 
\subsection{Finite Element Simulations} \label{s_simulations}
\par Simulations of cardiac mechanics were run in \textbf{C}Heart, a multi-physics finite element solver \cite{lee_multiphysics_2016}. Each mesh was discretized with quadratic elements, generated in Simmetrix SimModeler software [Simmetrix, 10 Executive Park Drive, Clifton Park, NY]. Table \ref{tab:n_elements} identifies the number of elements, the number of nodes, and mesh sizes for each geometry. Each coupled simulation was run for 30 cycles using 36 computer cores on Great Lakes, the University of Michigan high-performance computing core, whose standard partition consists of 455 computing nodes each comprised of 36 cores ( 3.0 GHz Intel Xeon Gold 6154 processors). 
\begin{table}[h]
\centering
\begin{tabular}{|p{9.5cm}|p{1.5cm}|p{1.5cm}|p{1.5cm}| }
\hline
\textbf{Geometry} & \textbf{$N_x$} & \textbf{$N_e$} & \textbf{$h$ [mm]} \\
\hline
Baseline Biventricular Geometry & 112828 & 70880 & 1.87 \\ \hline
Chronically Thinned Biventricular Heart & 133963 & 84713 & 1.69 \\ \hline
Baseline Biventricular Geometry with Surface Patch & 225934 & 151575 & 1.29 \\ \hline
Chronically Thinned Biventricular Heart with Surface Patch & 211246 & 140499 & 1.29 \\ \hline
\end{tabular}
\caption{Number of nodes $N_x$, number of elements $N_e$, and mesh sizes $h$ for study geometries}
\label{tab:n_elements}
\end{table}

\section{Results} 

\subsection{Baseline Cardiac Mechanics}

\begin{figure}[H]
\centering
\includegraphics[width=\textwidth]{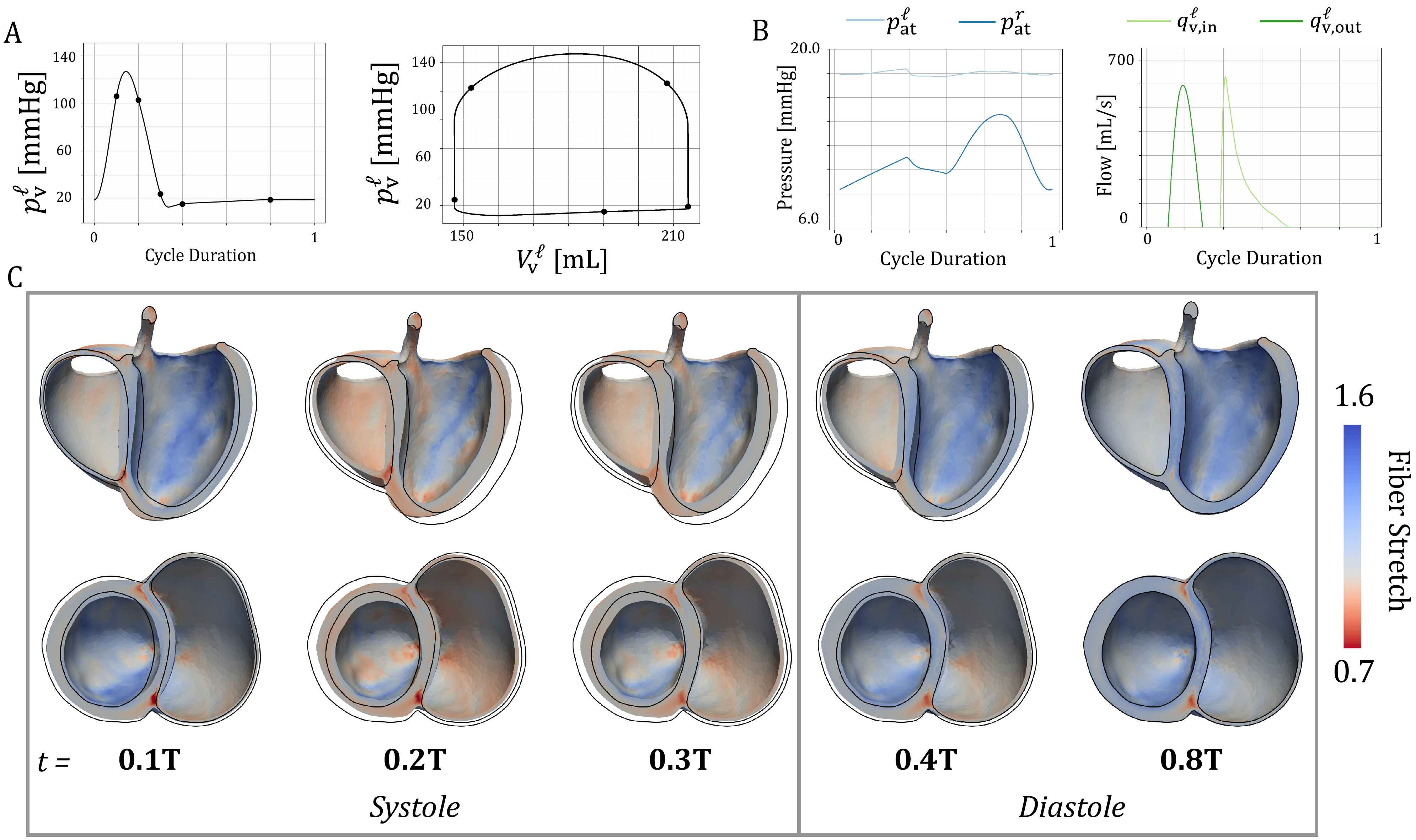}
\caption{Heart function for the healthy baseline heart model. A) LV pressure and LV pressure-volume relationship, with black dots indicating the illustrated times in Figure \ref{fig:healthy_fiber_stretch}C. B) 0D lumped parameter representation parameters for this model illustrated for one cardiac cycle, reflecting the simulated cardiovascular system - $p^l_{\text{at}}$ indicating left atrial pressure, $p^r_{\text{at}}$ indicating right atrial pressure, $q^l_{\text{v,in}}$ indicating mitral valve flow rate, and $q^l_{\text{v,out}}$ indicating aortic valve flow rate out of the LV. C) Fiber stretch illustrated at selected times in the cardiac cycle, with a short axis and long axis view. }
\label{fig:healthy_fiber_stretch}
\end{figure}

\par Figure \ref{fig:healthy_fiber_stretch} illustrates results from a baseline heart model with no fibrotic region. In Figure \ref{fig:healthy_fiber_stretch}A, the LV pressure and pressure-volume relationships are both illustrated, along with black dots indicating key times in the cardiac cycle. In Figure \ref{fig:healthy_fiber_stretch}B atrial pressures and flow into and out of the LV in the healthy case are shown. In Figure \ref{fig:healthy_fiber_stretch}C the fiber stretch is represented for each time point in the heart cycle as identified in Figure \ref{fig:healthy_fiber_stretch}A.  

\subsection{Transmural Patch}
 
\begin{figure}[H]
\centering
\includegraphics[width=\textwidth]{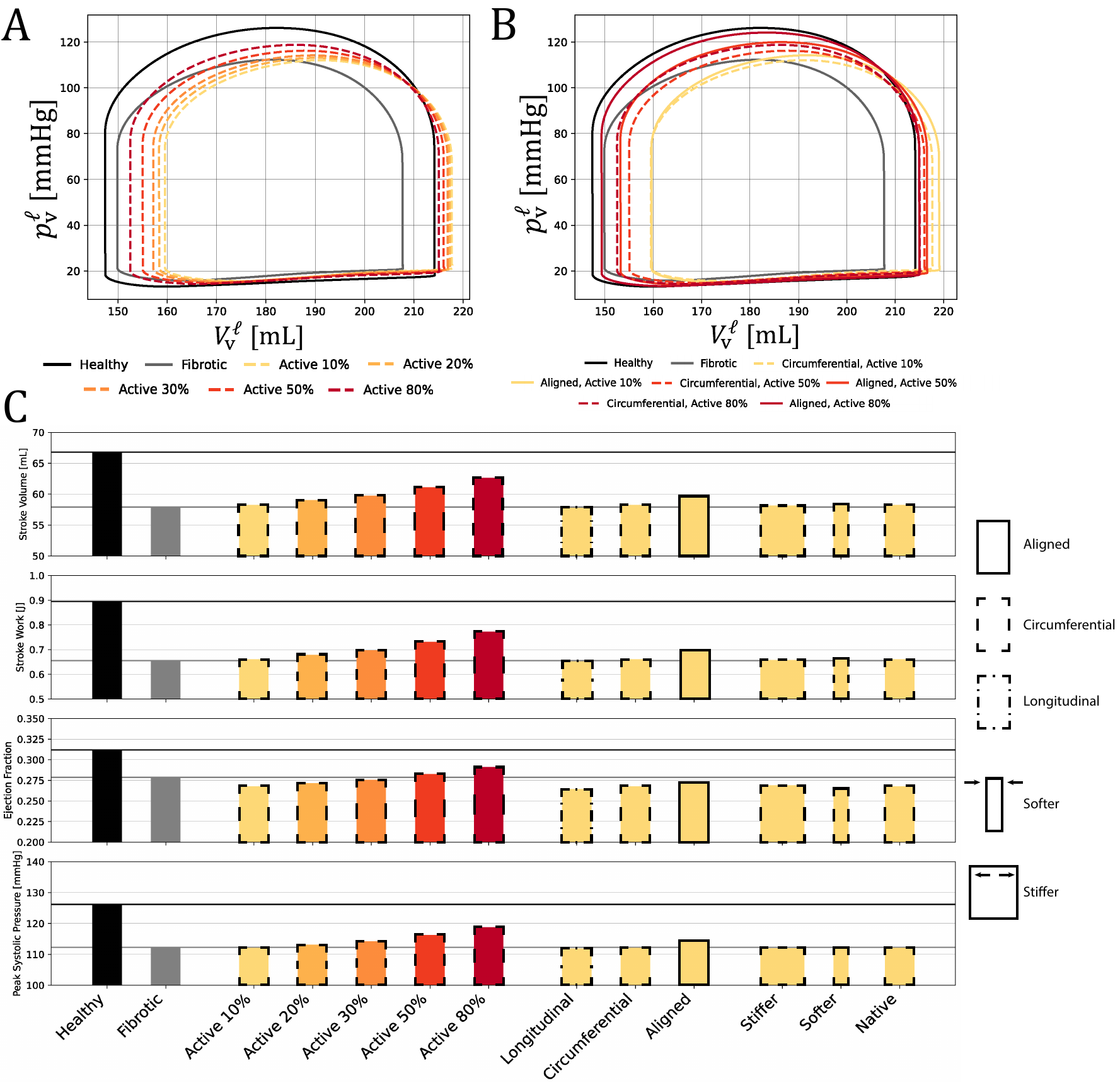}
\caption{Results for transmural patch variations. A) Pressure-Volume relationships for transmural patches with increasing activation, circumferential fiber orientation. B) Pressure-Volume relationships for transmural patches with different cardiac muscle fiber alignments and activation levels. C) Cardiac functional metrics for the transmural simulations, namely stroke volume, stroke work, ejection fraction, and peak systolic pressure. }
\label{fig:full_results}
\end{figure}

\par Figure \ref{fig:full_results}A shows the pressure-volume relationship of the LV of the heart as activation increases in the transmural cardiac tissue patch. In Figure \ref{fig:full_results}A, we highlight the impact of activation with circumferential cardiac muscle fibers. Figure \ref{fig:full_results}B shows the pressure-volume relationship of the LV of the heart for both circumferential and aligned cardiac muscle fibers. This illustrates the impact of aligning cardiac muscle fibers for different amounts of activation. Figure \ref{fig:full_results}C shows cardiac functional metrics - stroke volume, stroke work, ejection fraction, and peak systolic pressure. It shows the result of activation, fiber alignment, and stiffness from left to right. In this case, the activation simulations have circumferential fibers and native stiffness, the fiber alignment simulations have 10\% activation and native stiffness, and the stiffness simulations have circumferential fibers and 10\% activation with stiffer and softer tissue.

\subsection{Surface Patch}

\begin{figure}[H]
\centering
\includegraphics[width=\textwidth]{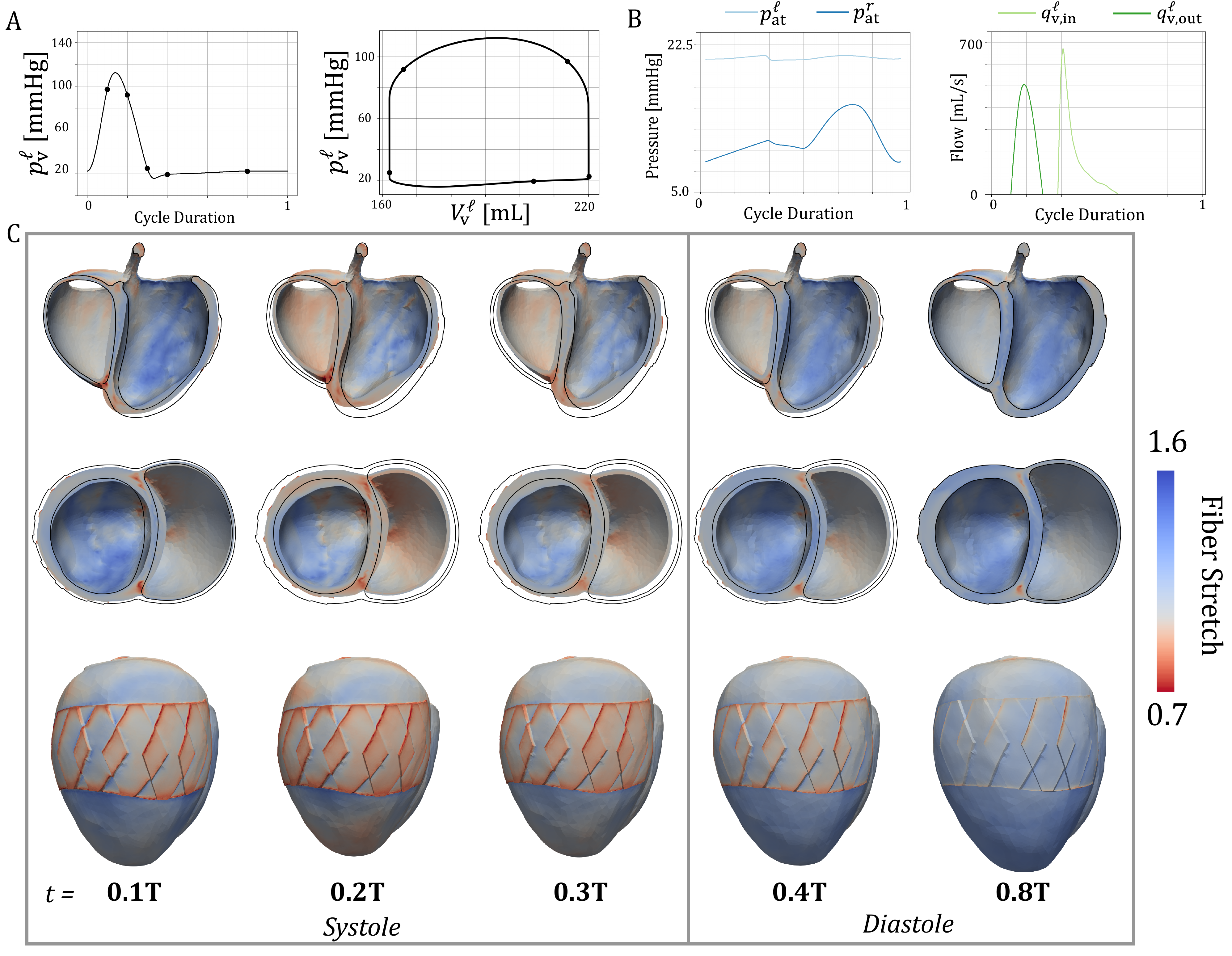}
\caption{Heart function for the surface heart model. A) LV pressure and LV pressure-volume relationship, with black dots indicating the illustrated times in Figure \ref{fig:healthy_fiber_stretch}C. B) 0D lumped parameter representation parameters for this model illustrated for one cardiac cycle - $p^l_{\text{at}}$ indicating left atrial pressure, $p^r_{\text{at}}$ indicating right atrial pressure, $q^l_{\text{v,in}}$ indicating mitral valve flow, and $q^l_{\text{v,out}}$ indicating aortic valve flow out of the LV. C) Fiber stretch illustrated at selected times in the cardiac cycle, with a short axis, a long axis view, and an epicardial view of the surface patch contracting. }
\label{fig:surf_fib_stretch}
\end{figure}

\par Figure \ref{fig:surf_fib_stretch} illustrates results from a surface patch heart model, in this case shown with 150\% activation. In Figure \ref{fig:surf_fib_stretch}A, the LV pressure and pressure-volume relationships are both illustrated, along with black dots indicating key times in the cardiac cycle. In Figure \ref{fig:surf_fib_stretch}B atrial pressures and flow into and out of the LV in the healthy baseline case are shown. In Figure \ref{fig:surf_fib_stretch}C the fiber stretch is represented for each time point in the heart cycle identified in Figure \ref{fig:surf_fib_stretch}A, including a view of the contracting surface patch.  

\begin{figure}[H]
\centering
\includegraphics[width=\textwidth]{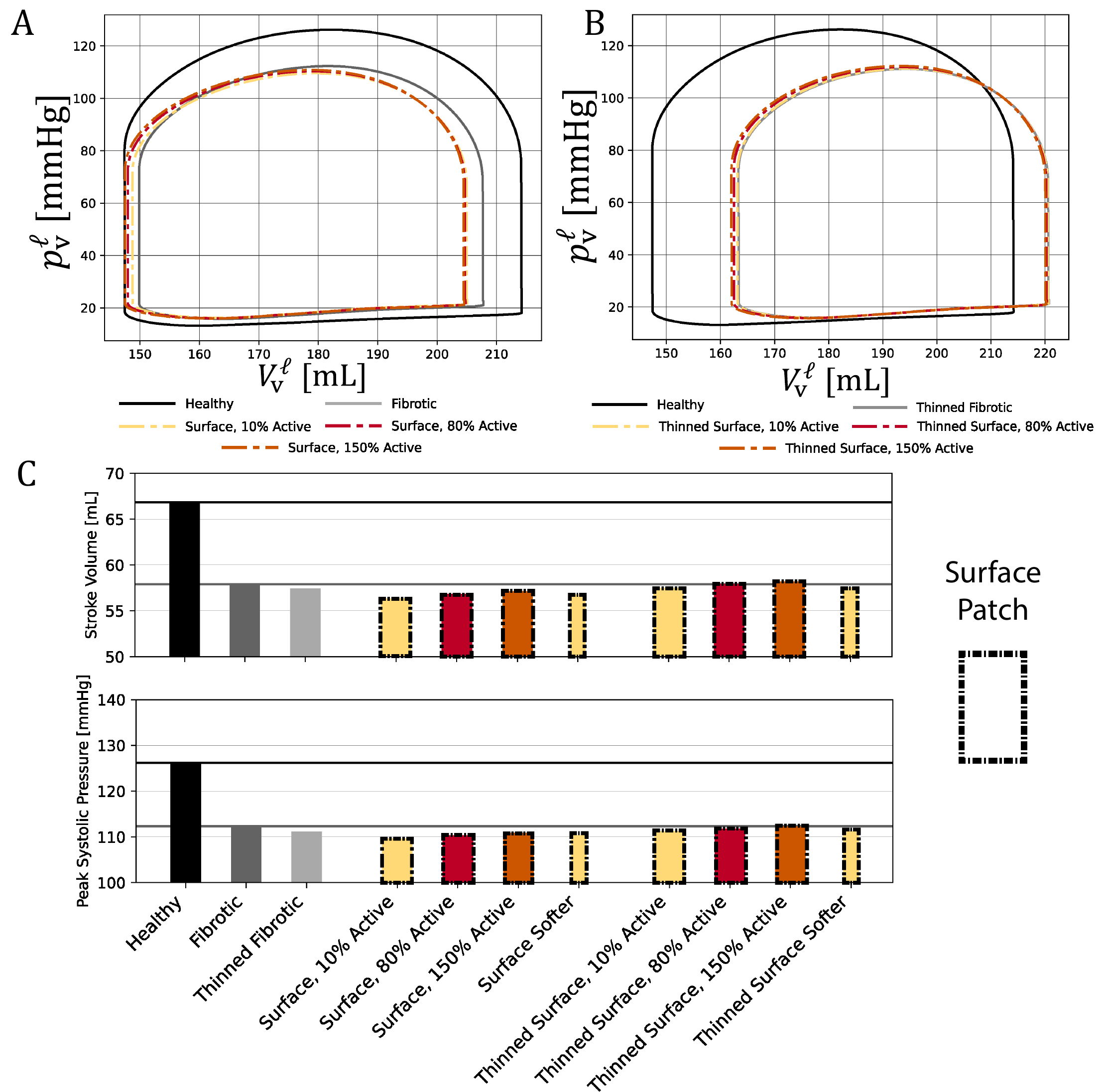}
\caption{Results for surface patch variations. A) Pressure-Volume relationships for surface patches with increasing activation with acute fibrosis. B) Pressure-Volume relationships for surface patches with increasing activation and chronically thinned fibrosis. C) Cardiac functional metrics for the surface patch simulations, namely stroke volume, stroke work, ejection fraction, and peak systolic pressure. }
\label{fig:surf_results}
\end{figure}

\par Figure \ref{fig:surf_results}A shows the pressure-volume relationship of the LV of the heart as activation increases in the surface cardiac tissue patch with an acutely fibrotic left ventricular morphology. Figure \ref{fig:surf_results}B shows the pressure-volume relationship of the LV of the heart as activation increases in the surface cardiac tissue patch with a chronically thinned fibrotic left ventricular morphology. Figure \ref{fig:surf_results}C shows cardiac functional metrics - stroke volume, stroke work, ejection fraction, and peak systolic pressure. It shows the results for the acutely fibrotic and chronically thinned fibrotic biventricular morphologies from left to right.

\subsection{Fiber Stretch}

\begin{figure}[H]
\centering
\includegraphics[width=\textwidth]{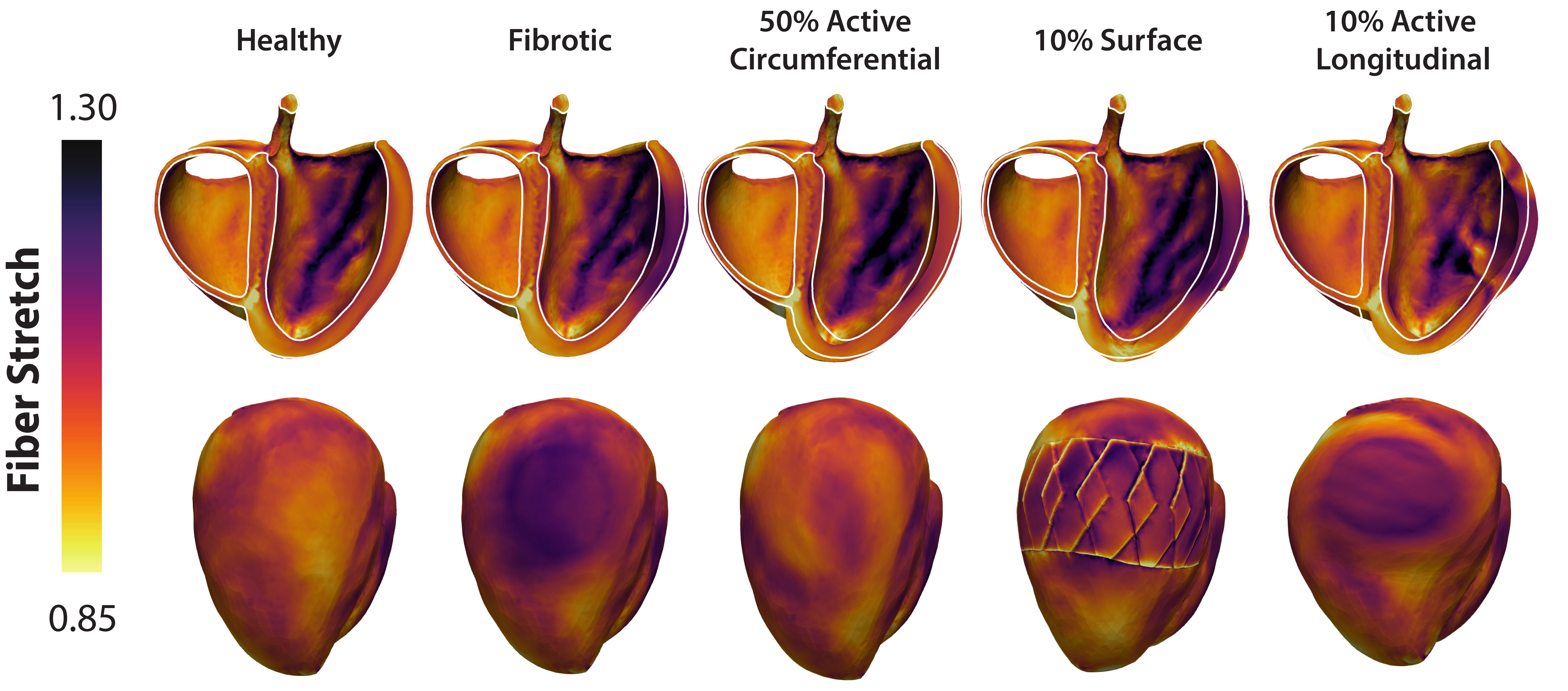}
\caption{Fiber stretch at $t=0.2T$ for selected transmural and surface patch variations. In the above biventricular view, the shape of the healthy baseline heart is traced in light gray.}
\label{fig:mech_diff_sys}
\end{figure}

Figure \ref{fig:mech_diff_sys} illustrates fiber stretch during systole for the healthy baseline case, the fibrotic case, the transmural patch with 50\% activation and circumferential fibers, the surface patch with 10\% activation, and the transmural patch case with 10\% active longitudinal fibers. The image illustrates how the fibrotic area of the LV contracts worse than the healthy baseline case, and the varying abilities or weaknesses of selected patches. A summary of all functional parameters for the patch simulations is shown in Table \ref{tab:cardiac_function_trans} in the Appendix.

\section{Discussion}

\par In this work, we modeled the impact of cardiac tissue-engineered patches to understand the implications of patch properties for patient heart function. We incorporated a solid mechanical model for cardiac tissue that reflects the microstructure of the heart. We simulated the influence of a fibrotic region, and then modeled both transmural and surface cardiac tissue patches with varying structural properties reflecting degrees of maturity. By meaningfully sweeping the tissue patch parameter space, we not only illustrated the impact of patch structural properties such as activation, cardiac muscle fiber alignment, and material properties, but also compared patch function in different patch types and biventricular morphologies.



\subsection{Impact of Structural Patch Properties}


\subsubsection{Activation}
\par 
In this study, we altered the contractile properties of cardiac tissue in a transmural tissue-engineered patch. 
Changing the contractile properties allowed us to study how the developed active stress in the cardiac tissue patch impacts pump function. For tissue engineers, this change in active stress for a given contractile signal approximates an increase in cardiac maturity \cite{tenreiro_next_2021}. 
We found as expected that contraction is able to improve cardiac function in general. Figure \ref{fig:full_results}A shows how increasing the generation of active stress results in better cardiac functional output overall. 
The stroke work of the LV increases from $0.6556 \text{mJ}$ to $0.6605 \text{mJ}$ for a weakly activating transmural patch and jumps up to $0.6786 \text{mJ}$ with 20\% activation. The stroke work steadily increases as activation increases for both circumferential and native muscle fiber orientations. This indicates that more mature cardiomyocytes are expected to improve overall performance in both cardiac muscle fiber orientations. 
\par To put parameter values in context, we identify "fractional recovery" of a cardiac metric as the \% change of the metric interpolated from the comparable fibrotic case to the healthy baseline case. 

\begin{equation}
    EF_{FR_x}=\frac{EF_x - EF_f}{EF_h - EF_f}
\end{equation}

\par In this case, $EF_{FR_x}$ indicates our fractional recovery metric, $EF_x$ represents the ejection fraction for the patch case we study, $EF_h$ represents healthy baseline ejection fraction, and $EF_f$ represents fibrotic ejection fraction. For stroke work in transmural patches, we first see a fairly low 2\% fractional recovery with 10\% activation and circumferential fibers, jumping to 9.6\% for 20\% activation, and then steadily increasing until we arrive at 48.9\% fractional recovery for 80\% activation in the patch with circumferential fibers. This is consistent with the literature where increased cardiac tissue patch maturity, developed through the inclusion of cardiac fibroblasts, is found to improve heart function in a mouse model of MI, although the study employs the surface patch attachment method \cite{lou_cardiac_2023}. 
\par 
For the surface patch, cardiac tissue activation was also varied. 
We found that for a given increase in activation in the surface patch, the cardiac functional output does not improve as much as in the transmural patch. Comparing Figure \ref{fig:full_results}C and Figure \ref{fig:surf_results}C illustrates that the transmural patch is more effective than a surface patch for the same generated active stress. 
This is expected because the amount of active tissue in the transmural patch is higher than in the surface patch. Furthermore, the contractile transmural patch replaces fibrotic tissue rather than resisting its stiffness. In addition, stroke volume fractional recovery of the surface patch at 150\% activation is 7.5\%, indicating that even exaggerated cardiac tissue activation cannot greatly improve function in this attachment method, as we can see by comparing plots in Figure \ref{fig:fractional_recovery}. For the thinned surface patch, we compare its function with the function of the thinned fibrotic heart. 
Despite the relative difficulty for cardiac surface patches in our mechanical simulations, experimentation has shown that these patches are effective in improving cardiac pump function in mouse and macaque models \cite{jebran_engineered_2025, lou_cardiac_2023}. Additionally, there is some evidence of beneficial LV remodeling in a human subject \cite{jebran_engineered_2025}. The findings of this paper suggest then that cardiac functional improvement can be significantly attributed to beneficial growth and remodeling of the left ventricle following cardiac tissue patch implantation. Notably, a chronically thinned wall improves the effectiveness of activation in the case of a surface patch (Figure \ref{fig:surf_results}C), which is more representative of typical disease progression.  

\subsubsection{Fiber Alignment}

\par We investigated the result of cardiac muscle fiber alignment in the transmural cardiac tissue patch, first considering how circumferential fibers and longitudinal fibers compare to the native fiber orientation. 
We constructed a rule-based fiber orientation for the native heart, and modified the fiber direction in the transmural patch. This allowed us to simulate what an oriented patch with fibers in the circumferential or longitudinal directions would look like. 
We found that for the same amount of activation, native fibers work best to deliver pump function to the heart, followed by circumferential fibers, and the least effective orientation is longitudinal. 
This is reflected by the fractional recovery of stroke work, where native fibers generate 16\% fractional recovery, circumferential fibers generate 2\% fractional recovery, and longitudinal fibers make stroke work slightly worse than the fibrotic case. 
Figure \ref{fig:full_results}A illustrates that for a tissue engineering protocol that grants cardiomyocytes of a certain maturity, matching the native fiber direction is a worthwhile goal. 
\par
\begin{figure}[H]
\centering
\includegraphics[width=\textwidth]{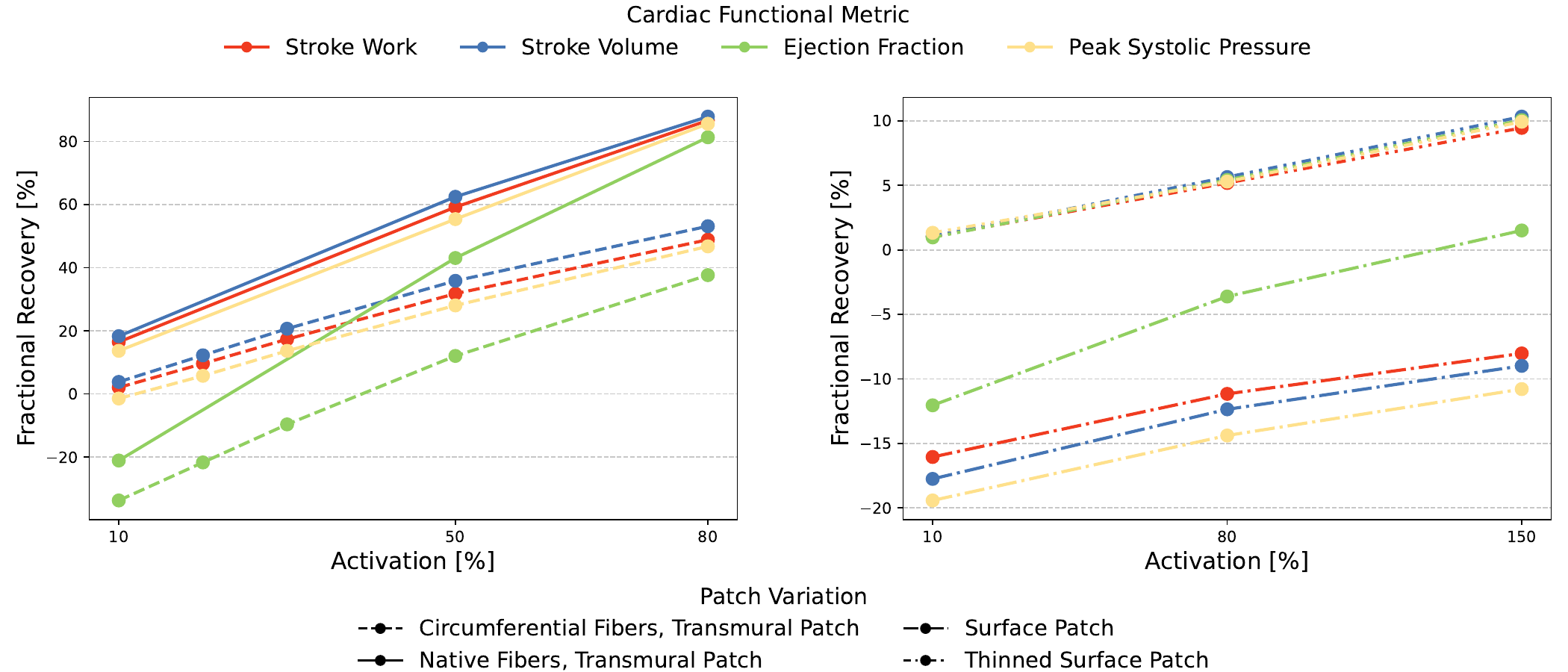}
\caption{Fractional Recovery of Cardiac Functional Metrics as a function of Activation for transmural and surface patch cases. In this case, patch variation is identified by line style and functional metric is identified by color. Note that both x-axes and y-axes differ between plots.}
\label{fig:fractional_recovery}
\end{figure}

 Alignment of cardiac muscle fibers in transmural tissue patches is modeled for different amounts of tissue activation. 
We examine these properties together to understand the improvement of heart pump function through muscle fiber alignment as generated active stress increases. 
We find that higher active stress increases the relative functional benefit of incorporating native cardiac muscle fibers in transmural cardiac tissue patches. Figure \ref{fig:full_results}B indicates that as active stress in the transmural patch goes up, functional improvement due to alignment of cardiac muscle fibers increases, which can also be seen in Figure \ref{fig:fractional_recovery}. 
For stroke work, the fractional recovery of a patch with 50\% activation and aligned muscle fibers is 59.2\%, while 80\% activation with circumferential fibers has a fractional recovery of only 48.9\% as seen in Figure \ref{fig:fractional_recovery}. 
Higher active stress in the transmural patch for circumferential fibers improves the gains we see as those fibers move to the native orientations, identifying native alignment as a valuable tool, especially as more mature cardiac tissue is developed. 
Figure \ref{fig:fractional_recovery} highlights the change in functional improvement between circumferential and aligned fibers as generated active stress increases.
\subsubsection{Tissue Stiffness}
%
\par We additionally examined the influence of cardiac tissue patch stiffness on heart function in both transmural and surface cardiac tissue patch attachment methods. 
We investigated stiffness because less mature cardiac tissue is known to be mechanically softer than mature tissue, and this may be an alterable property in engineered cardiac tissue patches \cite{kaiser_optimizing_2019}. 
In a transmural tissue patch, we found that a tissue patch that is softer than native tissue has slight benefits for the same amount of active stress and the same fiber direction. A transmural tissue patch that is stiffer than native tissue performs worse in terms of cardiac functional metrics, which we attribute to a lower fiber stretch for the same active stress. 
For the transmural application of the patch with 10\% activation and circumferential fibers, we see that stroke volume improvement relative to the fibrotic case is calculated to be 3.8\% with tissue of native stiffness, jumps to 5.3\% for softer tissue, and drops to 2.7\% for the stiffer tissue. Conversely, stroke volume in the epicardial surface patch decreases when material parameters soften, from 57.43 [$\textrm{mL}$] to 57.41 [$\textrm{mL}$].  
Consequently, tissue engineers should ensure that as cardiac tissue patches gain functional maturity, that they remain approximately as stiff as native tissue. For transmural applications, a softer patch may be beneficial, but more investigation is necessary to verify that the patch can handle cardiac pressures without rupture.  
\subsection{Patch Attachment Type and Left Ventricular Morphology} 
%
%
\par In this work, we investigated the effect of a chronically thinned left ventricular morphology on the function of a surface patch, as compared to a thicker region of fibrosis. 
We examined this because we expected the effectiveness of surface patch contraction to vary based on the left ventricular morphology. 
We find that stroke volume can better increase in the thinned case than in the acute thickened case, where the 150\% active surface patch on an acute left ventricular morphology decreases stroke volume relative to the fibrotic case by 12.4\% and the 150\% active surface patch on the thinned left ventricular morphology improves stroke volume by 5.6\%. Figure \ref{fig:surf_results}B indicates how the surface patch best improves pump function to a patient after fibrotic thinning has occurred. 
Adding a large volume of contractile tissue to the epicardium of a patient's heart is geometrically challenging, thus reflecting an advantage for surface patches applied after fibrotic thinning - the less fibrotic material for the surface patch to work against, the better this intervention may improve pump function. 
This combined with an easier surgical application identifies the surface patch as a strong option for older and more diseased patients, since the region may be further thinned, depending on how this remodeling impacts the stiffness of the fibrotic region. 
In previous work, researchers used an idealized left ventricle to study the impact of fibrotic stiffening and thinning on surface patch function \cite{janssens_role_2025}. It was similarly determined that when the left ventricular morphology is thinner, pump function improvement due to the cardiac tissue patch increases. 
Literature also indicates that aging is a significant risk factor for myocardial infarction generally, with different aging phenomena contributing to increased risk of cardiac fibrosis and worsening heart function \cite{shih_aging_2010, sanada_source_2018}. 
\subsection{Patch Attachment Strategy}
\par In this study, transmural and surface patch attachment methods are modeled to analyze the differences in performance between each strategy. 
We examined differences in performance to consider how cardiac tissue patch performance influences strategy selection. 
We found that the transmural patch is more effective than a surface patch for the same generated active stress (Figure \ref{fig:full_results}C and Figure \ref{fig:surf_results}C). 
If we measure pump function via stroke volume, a transmural patch with 80\% activation can generate 53.1\% fractional recovery, while the surface patch recovers just 1\% of the stroke volume. 
The stiffness of the fibrotic region post-MI has a valuable role in preventing rupture, despite the fact that this region's stiffness forms a stout barrier to increased cardiac function \cite{shinde_fibroblasts_2014, van_den_borne_myocardial_2010}. 
In this study, we assume that the transmural tissue patch surgical procedure removes all fibrotic tissue and replaces it with patch material, constituting a substantial assumption about the surgical application of such a patch and the capabilities of such patches. 
To rigorously design a cardiac tissue patch tailored to a specific patient, it is valuable to match structural properties like tissue stiffness, cardiac muscle fiber orientation, and surgical geometry. This process would require detailed knowledge of the patient's heart tissue characteristics, which can vary substantially \cite{liu_current_2019, kolawole_characterizing_2025}. This likely means that tissue engineers must investigate how cardiac tissue patches change properties in response to physiological application, and focus on personalizing the least flexible properties as best they are able \cite{jebran_engineered_2025}. 
\par Comparing the two patch application strategies reveals additional factors to consider. 
In both transmural and surface patches, doctors must suture the tissue-engineered construct to the native myocardium, which highlights potential issues associated with each type of patch.
For the transmural patch, patch application is a serious concern - cutting into a patient's heart wall would be a very invasive open-heart procedure, and it is unclear how such a transmural patch may attach to the surrounding heart wall or how it will interface with healthy tissue. 
For the surface patch, the surgical procedure would be less difficult, but attaching the soft patch to a patient heart around the epicardium may remain a challenge, with pericardial removal a potentially deleterious option \cite{pfaller_importance_2019}. 
The time point within the cardiac cycle where the patch is attached is also potentially influential - while in our case the surface patch is incorporated at the identified reference state, the surface patch could in theory be sutured at an arrested end-diastolic state instead. 
This could have the benefit of minimizing the diastolic interference of the surface patch, but it could problematically limit the Frank-Starling mechanism in the patch unless the patch is pre-stretched \cite{delicce_physiology_2025}.
More information would be needed to know the best time in the cardiac cycle to add a tissue-engineered cardiac tissue patch to the epicardial surface.

\subsection{Study Limitations}

\par One limitation of our approach is that we examined the effects of the cardiac tissue patch for a single representative patient rather than a collection of patients. In this work, we are interested in characterizing potential design choices for cardiac tissue patches, where a single representative patient is best for comparing different patches to the same fibrotic geometry. Future studies should consider increasing the sample size of patient geometries to investigate how trends hold across different heart shapes and potential fibrotic regions. Investigating the influence of a cardiac tissue patch in one patient lays the groundwork for studies investigating patient-specific impacts, identifying hypotheses for the impact of cardiac tissue patch design choices in a myocardial infarction patient cohort. 

\par Cardiac remodeling in response to the implantation of stem-cell-derived cardiac tissue is a nascent area of research. This current work does not consider post-intervention remodeling, neither deleterious nor advantageous. Some evidence suggests that adding a surface patch may induce beneficial remodeling, resulting in a fibrotic region that behaves more like native cardiac tissue \cite{jebran_engineered_2025, lou_cardiac_2023}. If such beneficial remodeling occurs, this could augment the effectiveness of a surface patch in particular. Additionally, both the patient heart and tissue patch can remodel, so patch structural properties could hypothetically mature in response to the cyclically contractile environment of a patient heart.
More research, experimental research in particular, is needed to determine these effects as the technology develops, which will contextualize the presented results.

\section{Conclusions}

\par In this work, we modeled the impact of cardiac tissue-engineered patches to understand the impact of patch properties. We employed a microstructurally-driven continuum solid mechanical model and coupled this solid model to a 0D lumped parameter representation of the circulatory system. We examined patch properties such as transmural patch activation, transmural cardiac fiber alignment, and surface patch activation. In doing so, we found that transmural cardiac tissue patches better improve function than surface cardiac tissue patches. Activation is identified as a crucial parameter for both transmural and surface cardiac tissue patch attachment methods. Native cardiac muscle fiber orientation improves function over circumferential alignment, and this effect grows as cardiac tissue active stress generation increases. 
Additionally, it was determined that surface cardiac tissue patches can slightly enhance heart function despite limiting diastolic filling, especially when fibrotic thinning has occurred. These conclusions identify broad design goals for the engineering of cardiac tissue patches to improve heart function after MI.

\section{Acknowledgments}

\par This work was supported by the National Science Foundation [grant number DGE-2241144]; and supported in part by the NSF Engineering Research Center CELL-MET [grant number EEC-1647837]. This research was supported in part through computational resources and services provided by Advanced Research Computing at the University of Michigan, Ann Arbor.

\section{CRediT Statement}
\par \textbf{John Patrick Sayut Jr.:} Conceptualization, Methodology, Software, Investigation, Writing - Original Draft, Visualization \textbf{Javiera Jilberto:} Methodology, Software, Writing - Review \& Editing \textbf{Mia Bonini:} Methodology, Software, Data Curation \textbf{Marc Hirschvogel:} Methodology, Software, Validation, Data Curation \textbf{Will Zhang:} Methodology, Writing - Review \& Editing \textbf{David Nordsletten:} Conceptualization, Methodology, Software, Resources, Writing - Review \& Editing

\section{Ethics Statement}
\par The data collection procedure in this work was performed in compliance with relevant laws and institutional guidelines and have been approved by the appropriate institutional committees. More information can be found in IRB HUM00196629.

\bibliographystyle{elsarticle-num}
\bibliography{patchpapercite}

\begin{thebibliography}{10}
\expandafter\ifx\csname url\endcsname\relax
  \def\url#1{\texttt{#1}}\fi
\expandafter\ifx\csname urlprefix\endcsname\relax\def\urlprefix{URL }\fi
\expandafter\ifx\csname href\endcsname\relax
  \def\href#1#2{#2} \def\path#1{#1}\fi

\bibitem{tunstall-pedoe_myocardial_1994}
H.~Tunstall-Pedoe, K.~Kuulasmaa, P.~Amouyel, D.~Arveiler, A.~M. Rajakangas, A.~Pajak, \href{https://www.ahajournals.org/doi/10.1161/01.CIR.90.1.583}{Myocardial infarction and coronary deaths in the world health organization {MONICA} project. registration procedures, event rates, and case-fatality rates in 38 populations from 21 countries in four continents.} 90~(1)  583--612.
\newblock \href {https://doi.org/10.1161/01.CIR.90.1.583} {\path{doi:10.1161/01.CIR.90.1.583}}.
\newline\urlprefix\url{https://www.ahajournals.org/doi/10.1161/01.CIR.90.1.583}

\bibitem{hinderer_cardiac_2019}
S.~Hinderer, K.~Schenke-Layland, \href{https://linkinghub.elsevier.com/retrieve/pii/S0169409X19300614}{Cardiac fibrosis – a short review of causes and therapeutic strategies} 146  77--82.
\newblock \href {https://doi.org/10.1016/j.addr.2019.05.011} {\path{doi:10.1016/j.addr.2019.05.011}}.
\newline\urlprefix\url{https://linkinghub.elsevier.com/retrieve/pii/S0169409X19300614}

\bibitem{salari_global_2023}
N.~Salari, F.~Morddarvanjoghi, A.~Abdolmaleki, S.~Rasoulpoor, A.~A. Khaleghi, L.~A. Hezarkhani, S.~Shohaimi, M.~Mohammadi, \href{https://doi.org/10.1186/s12872-023-03231-w}{The global prevalence of myocardial infarction: a systematic review and meta-analysis} 23~(1)  206.
\newblock \href {https://doi.org/10.1186/s12872-023-03231-w} {\path{doi:10.1186/s12872-023-03231-w}}.
\newline\urlprefix\url{https://doi.org/10.1186/s12872-023-03231-w}

\bibitem{cameli_donor_2022}
M.~Cameli, M.~C. Pastore, A.~Campora, M.~Lisi, G.~E. Mandoli, \href{https://www.ncbi.nlm.nih.gov/pmc/articles/PMC9618685/}{Donor shortage in heart transplantation: How can we overcome this challenge?} 9  1001002.
\newblock \href {https://doi.org/10.3389/fcvm.2022.1001002} {\path{doi:10.3389/fcvm.2022.1001002}}.
\newline\urlprefix\url{https://www.ncbi.nlm.nih.gov/pmc/articles/PMC9618685/}

\bibitem{estep_risk_2015}
J.~D. Estep, R.~C. Starling, D.~A. Horstmanshof, C.~A. Milano, C.~H. Selzman, K.~B. Shah, M.~Loebe, N.~Moazami, J.~W. Long, J.~Stehlik, V.~Kasirajan, D.~C. Haas, J.~B. O'Connell, A.~J. Boyle, D.~J. Farrar, J.~G. Rogers, \href{https://linkinghub.elsevier.com/retrieve/pii/S0735109715048779}{Risk assessment and comparative effectiveness of left ventricular assist device and medical management in ambulatory heart failure patients} 66~(16)  1747--1761.
\newblock \href {https://doi.org/10.1016/j.jacc.2015.07.075} {\path{doi:10.1016/j.jacc.2015.07.075}}.
\newline\urlprefix\url{https://linkinghub.elsevier.com/retrieve/pii/S0735109715048779}

\bibitem{llerena-velastegui_efficacy_2024}
J.~Llerena-Velastegui, G.~Santafe-Abril, C.~Villacis-Lopez, C.~Hurtado-Alzate, M.~Placencia-Silva, M.~Santander-Aldean, M.~Trujillo-Delgado, X.~Freire-Oña, C.~Santander-Fuentes, J.~Velasquez-Campos, \href{https://linkinghub.elsevier.com/retrieve/pii/S0146280623005352}{Efficacy and complication profiles of left ventricular assist devices in adult heart failure management: A systematic review and meta-analysis} 49~(1)  102118.
\newblock \href {https://doi.org/10.1016/j.cpcardiol.2023.102118} {\path{doi:10.1016/j.cpcardiol.2023.102118}}.
\newline\urlprefix\url{https://linkinghub.elsevier.com/retrieve/pii/S0146280623005352}

\bibitem{tenreiro_next_2021}
M.~F. Tenreiro, A.~F. Louro, P.~M. Alves, M.~Serra, \href{https://www.nature.com/articles/s41536-021-00140-4}{Next generation of heart regenerative therapies: progress and promise of cardiac tissue engineering} 6~(1)  30.
\newblock \href {https://doi.org/10.1038/s41536-021-00140-4} {\path{doi:10.1038/s41536-021-00140-4}}.
\newline\urlprefix\url{https://www.nature.com/articles/s41536-021-00140-4}

\bibitem{liu_human_2018}
Y.-W. Liu, B.~Chen, X.~Yang, J.~A. Fugate, F.~A. Kalucki, A.~Futakuchi-Tsuchida, L.~Couture, K.~W. Vogel, C.~A. Astley, A.~Baldessari, J.~Ogle, C.~W. Don, Z.~L. Steinberg, S.~P. Seslar, S.~A. Tuck, H.~Tsuchida, A.~V. Naumova, S.~K. Dupras, M.~S. Lyu, J.~Lee, D.~W. Hailey, H.~Reinecke, L.~Pabon, B.~H. Fryer, W.~R. {MacLellan}, R.~S. Thies, C.~E. Murry, \href{https://www.nature.com/articles/nbt.4162}{Human embryonic stem cell–derived cardiomyocytes restore function in infarcted hearts of non-human primates} 36~(7)  597--605.
\newblock \href {https://doi.org/10.1038/nbt.4162} {\path{doi:10.1038/nbt.4162}}.
\newline\urlprefix\url{https://www.nature.com/articles/nbt.4162}

\bibitem{marchiano_gene_2023}
S.~Marchiano, K.~Nakamura, H.~Reinecke, L.~Neidig, M.~Lai, S.~Kadota, F.~Perbellini, X.~Yang, J.~M. Klaiman, L.~P. Blakely, E.~Karbassi, P.~A. Fields, A.~M. Fenix, K.~M. Beussman, A.~Jayabalu, F.~A. Kalucki, J.~C. Potter, A.~Futakuchi-Tsuchida, G.~J. Weber, S.~Dupras, H.~Tsuchida, L.~Pabon, L.~Wang, B.~C. Knollmann, S.~Kattman, R.~S. Thies, N.~Sniadecki, W.~R. {MacLellan}, A.~Bertero, C.~E. Murry, \href{https://linkinghub.elsevier.com/retrieve/pii/S1934590923000814}{Gene editing to prevent ventricular arrhythmias associated with cardiomyocyte cell therapy} 30~(4)  396--414.e9.
\newblock \href {https://doi.org/10.1016/j.stem.2023.03.010} {\path{doi:10.1016/j.stem.2023.03.010}}.
\newline\urlprefix\url{https://linkinghub.elsevier.com/retrieve/pii/S1934590923000814}

\bibitem{nakamura_pharmacologic_2021}
K.~Nakamura, L.~E. Neidig, X.~Yang, G.~J. Weber, D.~El-Nachef, H.~Tsuchida, S.~Dupras, F.~A. Kalucki, A.~Jayabalu, A.~Futakuchi-Tsuchida, D.~S. Nakamura, S.~Marchianò, A.~Bertero, M.~R. Robinson, K.~Cain, D.~Whittington, R.~Tian, H.~Reinecke, L.~Pabon, B.~C. Knollmann, S.~Kattman, R.~S. Thies, W.~R. {MacLellan}, C.~E. Murry, \href{https://linkinghub.elsevier.com/retrieve/pii/S2213671121004239}{Pharmacologic therapy for engraftment arrhythmia induced by transplantation of human cardiomyocytes} 16~(10)  2473--2487.
\newblock \href {https://doi.org/10.1016/j.stemcr.2021.08.005} {\path{doi:10.1016/j.stemcr.2021.08.005}}.
\newline\urlprefix\url{https://linkinghub.elsevier.com/retrieve/pii/S2213671121004239}

\bibitem{tsan_physiologic_2021}
Y.-C. Tsan, S.~J. {DePalma}, Y.-T. Zhao, A.~Capilnasiu, Y.-W. Wu, B.~Elder, I.~Panse, K.~Ufford, D.~L. Matera, S.~Friedline, T.~S. O’Leary, N.~Wubshet, K.~K.~Y. Ho, M.~J. Previs, D.~Nordsletten, L.~L. Isom, B.~M. Baker, A.~P. Liu, A.~S. Helms, \href{https://www.nature.com/articles/s41467-021-26496-1}{Physiologic biomechanics enhance reproducible contractile development in a stem cell derived cardiac muscle platform} 12~(1)  6167, publisher: Nature Publishing Group.
\newblock \href {https://doi.org/10.1038/s41467-021-26496-1} {\path{doi:10.1038/s41467-021-26496-1}}.
\newline\urlprefix\url{https://www.nature.com/articles/s41467-021-26496-1}

\bibitem{shadrin_cardiopatch_2017}
I.~Y. Shadrin, B.~W. Allen, Y.~Qian, C.~P. Jackman, A.~L. Carlson, M.~E. Juhas, N.~Bursac, \href{https://www.nature.com/articles/s41467-017-01946-x}{Cardiopatch platform enables maturation and scale-up of human pluripotent stem cell-derived engineered heart tissues} 8~(1)  1825.
\newblock \href {https://doi.org/10.1038/s41467-017-01946-x} {\path{doi:10.1038/s41467-017-01946-x}}.
\newline\urlprefix\url{https://www.nature.com/articles/s41467-017-01946-x}

\bibitem{jebran_engineered_2025}
A.-F. Jebran, T.~Seidler, M.~Tiburcy, M.~Daskalaki, I.~Kutschka, B.~Fujita, S.~Ensminger, F.~Bremmer, A.~Moussavi, H.~Yang, X.~Qin, S.~Mißbach, C.~Drummer, H.~Baraki, S.~Boretius, C.~Hasenauer, T.~Nette, J.~Kowallick, C.~O. Ritter, J.~Lotz, M.~Didié, M.~Mietsch, T.~Meyer, G.~Kensah, D.~Krüger, M.~S. Sakib, L.~Kaurani, A.~Fischer, R.~Dressel, I.~Rodriguez-Polo, M.~Stauske, S.~Diecke, K.~Maetz-Rensing, E.~Gruber-Dujardin, M.~Bleyer, B.~Petersen, C.~Roos, L.~Zhang, L.~Walter, S.~Kaulfuß, G.~Yigit, B.~Wollnik, E.~Levent, B.~Roshani, C.~Stahl-Henning, P.~Ströbel, T.~Legler, J.~Riggert, K.~Hellenkamp, J.-U. Voigt, G.~Hasenfuß, R.~Hinkel, J.~C. Wu, R.~Behr, W.-H. Zimmermann, \href{https://www.nature.com/articles/s41586-024-08463-0}{Engineered heart muscle allografts for heart repair in primates and humans}  1--9Publisher: Nature Publishing Group.
\newblock \href {https://doi.org/10.1038/s41586-024-08463-0} {\path{doi:10.1038/s41586-024-08463-0}}.
\newline\urlprefix\url{https://www.nature.com/articles/s41586-024-08463-0}

\bibitem{zimmermann_multilayer_2023}
W.-H. Zimmermann, M.~{TIBURCY}, T.~Meyer, \href{https://patents.google.com/patent/US20230390460A1/en}{Multilayer engineered heart muscle}.
\newline\urlprefix\url{https://patents.google.com/patent/US20230390460A1/en}

\bibitem{jabbour_vivo_2021}
R.~J. Jabbour, T.~J. Owen, P.~Pandey, M.~Reinsch, B.~Wang, O.~King, L.~S. Couch, D.~Pantou, D.~S. Pitcher, R.~A. Chowdhury, F.~G. Pitoulis, B.~S. Handa, W.~Kit-Anan, F.~Perbellini, R.~C. Myles, D.~J. Stuckey, M.~Dunne, M.~Shanmuganathan, N.~S. Peters, F.~S. Ng, F.~Weinberger, C.~M. Terracciano, G.~L. Smith, T.~Eschenhagen, S.~E. Harding, \href{https://insight.jci.org/articles/view/144068}{In vivo grafting of large engineered heart tissue patches for cardiac repair} 6~(15)  e144068.
\newblock \href {https://doi.org/10.1172/jci.insight.144068} {\path{doi:10.1172/jci.insight.144068}}.
\newline\urlprefix\url{https://insight.jci.org/articles/view/144068}

\bibitem{guan_transplantation_2020}
X.~Guan, W.~Xu, H.~Zhang, Q.~Wang, J.~Yu, R.~Zhang, Y.~Chen, Y.~Xia, J.~Wang, D.~Wang, \href{https://stemcellres.biomedcentral.com/articles/10.1186/s13287-020-01602-0}{Transplantation of human induced pluripotent stem cell-derived cardiomyocytes improves myocardial function and reverses ventricular remodeling in infarcted rat hearts} 11~(1)  73.
\newblock \href {https://doi.org/10.1186/s13287-020-01602-0} {\path{doi:10.1186/s13287-020-01602-0}}.
\newline\urlprefix\url{https://stemcellres.biomedcentral.com/articles/10.1186/s13287-020-01602-0}

\bibitem{lou_cardiac_2023}
X.~Lou, Y.~Tang, L.~Ye, D.~Pretorius, V.~G. Fast, A.~M. Kahn-Krell, J.~Zhang, J.~Zhang, A.~Qiao, G.~Qin, T.~Kamp, J.~A. Thomson, J.~Zhang, Cardiac muscle patches containing four types of cardiac cells derived from human pluripotent stem cells improve recovery from cardiac injury in mice 119~(4)  1062--1076.
\newblock \href {https://doi.org/10.1093/cvr/cvad004} {\path{doi:10.1093/cvr/cvad004}}.

\bibitem{gao_large_2018}
L.~Gao, Z.~R. Gregorich, W.~Zhu, S.~Mattapally, Y.~Oduk, X.~Lou, R.~Kannappan, A.~V. Borovjagin, G.~P. Walcott, A.~E. Pollard, V.~G. Fast, X.~Hu, S.~G. Lloyd, Y.~Ge, J.~Zhang, \href{https://www.ahajournals.org/doi/10.1161/CIRCULATIONAHA.117.030785}{Large cardiac muscle patches engineered from human induced-pluripotent stem cell–derived cardiac cells improve recovery from myocardial infarction in swine} 137~(16)  1712--1730.
\newblock \href {https://doi.org/10.1161/CIRCULATIONAHA.117.030785} {\path{doi:10.1161/CIRCULATIONAHA.117.030785}}.
\newline\urlprefix\url{https://www.ahajournals.org/doi/10.1161/CIRCULATIONAHA.117.030785}

\bibitem{syed_modeling_2023}
F.~Syed, S.~Khan, M.~Toma, \href{https://www.mdpi.com/2079-7737/12/7/1026}{Modeling dynamics of the cardiovascular system using fluid-structure interaction methods} 12~(7)  1026, number: 7 Publisher: Multidisciplinary Digital Publishing Institute.
\newblock \href {https://doi.org/10.3390/biology12071026} {\path{doi:10.3390/biology12071026}}.
\newline\urlprefix\url{https://www.mdpi.com/2079-7737/12/7/1026}

\bibitem{hirschvogel_monolithic_2017}
M.~Hirschvogel, M.~Bassilious, L.~Jagschies, S.~M. Wildhirt, M.~W. Gee, \href{https://onlinelibrary.wiley.com/doi/10.1002/cnm.2842}{A monolithic 3d-0d coupled closed-loop model of the heart and the vascular system: Experiment-based parameter estimation for patient-specific cardiac mechanics: 3d-0d coupled closed-loop model of the heart} 33~(8)  e2842.
\newblock \href {https://doi.org/10.1002/cnm.2842} {\path{doi:10.1002/cnm.2842}}.
\newline\urlprefix\url{https://onlinelibrary.wiley.com/doi/10.1002/cnm.2842}

\bibitem{miller_implementation_2021}
R.~Miller, E.~Kerfoot, C.~Mauger, T.~F. Ismail, A.~A. Young, D.~A. Nordsletten, \href{https://www.frontiersin.org/articles/10.3389/fphys.2021.716597/full}{An implementation of patient-specific biventricular mechanics simulations with a deep learning and computational pipeline} 12  716597.
\newblock \href {https://doi.org/10.3389/fphys.2021.716597} {\path{doi:10.3389/fphys.2021.716597}}.
\newline\urlprefix\url{https://www.frontiersin.org/articles/10.3389/fphys.2021.716597/full}

\bibitem{hirschvogel_silico_2019}
M.~Hirschvogel, L.~Jagschies, A.~Maier, S.~M. Wildhirt, M.~W. Gee, \href{https://onlinelibrary.wiley.com/doi/abs/10.1002/cnm.3233}{An in silico twin for epicardial augmentation of the failing heart} 35~(10)  e3233, \_eprint: https://onlinelibrary.wiley.com/doi/pdf/10.1002/cnm.3233.
\newblock \href {https://doi.org/10.1002/cnm.3233} {\path{doi:10.1002/cnm.3233}}.
\newline\urlprefix\url{https://onlinelibrary.wiley.com/doi/abs/10.1002/cnm.3233}

\bibitem{ohara_personalized_2022}
R.~P. O'Hara, E.~Binka, A.~Prakosa, S.~L. Zimmerman, M.~J. Cartoski, M.~R. Abraham, D.-Y. Lu, P.~M. Boyle, N.~A. Trayanova, \href{https://doi.org/10.7554/eLife.73325}{Personalized computational heart models with t1-mapped fibrotic remodeling predict sudden death risk in patients with hypertrophic cardiomyopathy} 11  e73325, publisher: {eLife} Sciences Publications, Ltd.
\newblock \href {https://doi.org/10.7554/eLife.73325} {\path{doi:10.7554/eLife.73325}}.
\newline\urlprefix\url{https://doi.org/10.7554/eLife.73325}

\bibitem{niederer_creation_2020}
S.~A. Niederer, Y.~Aboelkassem, C.~D. Cantwell, C.~Corrado, S.~Coveney, E.~M. Cherry, T.~Delhaas, F.~H. Fenton, A.~V. Panfilov, P.~Pathmanathan, G.~Plank, M.~Riabiz, C.~H. Roney, R.~W. dos Santos, L.~Wang, \href{https://royalsocietypublishing.org/doi/full/10.1098/rsta.2019.0558}{Creation and application of virtual patient cohorts of heart models} 378~(2173)  20190558, publisher: Royal Society.
\newblock \href {https://doi.org/10.1098/rsta.2019.0558} {\path{doi:10.1098/rsta.2019.0558}}.
\newline\urlprefix\url{https://royalsocietypublishing.org/doi/full/10.1098/rsta.2019.0558}

\bibitem{janssens_post-infarct_2023}
K.~L. P.~M. Janssens, M.~Kraamer, L.~Barbarotta, P.~H.~M. Bovendeerd, Post-infarct evolution of ventricular and myocardial function 22~(6)  1815--1828.
\newblock \href {https://doi.org/10.1007/s10237-023-01734-1} {\path{doi:10.1007/s10237-023-01734-1}}.

\bibitem{mojsejenko_estimating_2015}
D.~Mojsejenko, J.~R. {McGarvey}, S.~M. Dorsey, J.~H. Gorman, J.~A. Burdick, J.~J. Pilla, R.~C. Gorman, J.~F. Wenk, \href{http://link.springer.com/10.1007/s10237-014-0627-z}{Estimating passive mechanical properties in a myocardial infarction using {MRI} and finite element simulations} 14~(3)  633--647.
\newblock \href {https://doi.org/10.1007/s10237-014-0627-z} {\path{doi:10.1007/s10237-014-0627-z}}.
\newline\urlprefix\url{http://link.springer.com/10.1007/s10237-014-0627-z}

\bibitem{janssens_impact_2024}
K.~L. P.~M. Janssens, P.~H.~M. Bovendeerd, \href{https://link.springer.com/10.1007/s10237-024-01877-9}{Impact of cardiac patch alignment on restoring post-infarct ventricular function} 23~(6)  1963--1976.
\newblock \href {https://doi.org/10.1007/s10237-024-01877-9} {\path{doi:10.1007/s10237-024-01877-9}}.
\newline\urlprefix\url{https://link.springer.com/10.1007/s10237-024-01877-9}

\bibitem{janssens_role_2025}
K.~L. P.~M. Janssens, P.~H.~M. Bovendeerd, \href{https://doi.org/10.1115/1.4068829}{The role of infarct stiffness in cardiac patch therapy: A computational study using an idealized left ventricular geometry} 147~(81005).
\newblock \href {https://doi.org/10.1115/1.4068829} {\path{doi:10.1115/1.4068829}}.
\newline\urlprefix\url{https://doi.org/10.1115/1.4068829}

\bibitem{holzapfel_nonlinear_2010}
G.~A. Holzapfel, Nonlinear solid mechanics: a continuum approach for engineering, repr Edition, Wiley.

\bibitem{nordsletten_viscoelastic_2021}
D.~Nordsletten, A.~Capilnasiu, W.~Zhang, A.~Wittgenstein, M.~Hadjicharalambous, G.~Sommer, R.~Sinkus, G.~A. Holzapfel, \href{https://www.sciencedirect.com/science/article/pii/S1742706121005699}{A viscoelastic model for human myocardium} 135  441--457.
\newblock \href {https://doi.org/10.1016/j.actbio.2021.08.036} {\path{doi:10.1016/j.actbio.2021.08.036}}.
\newline\urlprefix\url{https://www.sciencedirect.com/science/article/pii/S1742706121005699}

\bibitem{kerckhoffs_homogeneity_2003}
R.~C.~P. Kerckhoffs, P.~H.~M. Bovendeerd, J.~C.~S. Kotte, F.~W. Prinzen, K.~Smits, T.~Arts, \href{http://link.springer.com/10.1114/1.1566447}{Homogeneity of cardiac contraction despite physiological asynchrony of depolarization: A model study} 31~(5)  536--547.
\newblock \href {https://doi.org/10.1114/1.1566447} {\path{doi:10.1114/1.1566447}}.
\newline\urlprefix\url{http://link.springer.com/10.1114/1.1566447}

\bibitem{tangney_novel_2013}
J.~Tangney, J.~Chuang, M.~Janssen, A.~Krishnamurthy, P.~Liao, M.~Hoshijima, X.~Wu, G.~Meininger, M.~Muthuchamy, A.~Zemljic-Harpf, R.~Ross, L.~Frank, A.~{McCulloch}, J.~Omens, \href{https://linkinghub.elsevier.com/retrieve/pii/S0006349513002385}{Novel role for vinculin in ventricular myocyte mechanics and dysfunction} 104~(7)  1623--1633.
\newblock \href {https://doi.org/10.1016/j.bpj.2013.02.021} {\path{doi:10.1016/j.bpj.2013.02.021}}.
\newline\urlprefix\url{https://linkinghub.elsevier.com/retrieve/pii/S0006349513002385}

\bibitem{kassab_microstructurally_2016}
A.~Krishnamurthy, B.~Coppola, J.~Tangney, R.~C.~P. Kerckhoffs, J.~H. Omens, A.~D. {McCulloch}, \href{https://link.springer.com/10.1007/978-1-4899-7630-7_22}{A microstructurally based multi-scale constitutive model of active myocardial mechanics}, in: G.~S. Kassab, M.~S. Sacks (Eds.), Structure-Based Mechanics of Tissues and Organs, Springer {US}, pp. 439--460.
\newblock \href {https://doi.org/10.1007/978-1-4899-7630-7_22} {\path{doi:10.1007/978-1-4899-7630-7_22}}.
\newline\urlprefix\url{https://link.springer.com/10.1007/978-1-4899-7630-7_22}

\bibitem{bayer_novel_2012}
J.~D. Bayer, R.~C. Blake, G.~Plank, N.~A. Trayanova, \href{https://doi.org/10.1007/s10439-012-0593-5}{A novel rule-based algorithm for assigning myocardial fiber orientation to computational heart models} 40~(10)  2243--2254.
\newblock \href {https://doi.org/10.1007/s10439-012-0593-5} {\path{doi:10.1007/s10439-012-0593-5}}.
\newline\urlprefix\url{https://doi.org/10.1007/s10439-012-0593-5}

\bibitem{doste_rule-based_2019}
R.~Doste, D.~Soto-Iglesias, G.~Bernardino, A.~Alcaine, R.~Sebastian, S.~Giffard-Roisin, M.~Sermesant, A.~Berruezo, D.~Sanchez-Quintana, O.~Camara, \href{https://onlinelibrary.wiley.com/doi/abs/10.1002/cnm.3185}{A rule-based method to model myocardial fiber orientation in cardiac biventricular geometries with outflow tracts} 35~(4)  e3185, \_eprint: https://onlinelibrary.wiley.com/doi/pdf/10.1002/cnm.3185.
\newblock \href {https://doi.org/10.1002/cnm.3185} {\path{doi:10.1002/cnm.3185}}.
\newline\urlprefix\url{https://onlinelibrary.wiley.com/doi/abs/10.1002/cnm.3185}

\bibitem{klotz_single-beat_2006}
S.~Klotz, I.~Hay, M.~L. Dickstein, G.-H. Yi, J.~Wang, M.~S. Maurer, D.~A. Kass, D.~Burkhoff, \href{https://journals.physiology.org/doi/full/10.1152/ajpheart.01240.2005}{Single-beat estimation of end-diastolic pressure-volume relationship: a novel method with potential for noninvasive application} 291~(1)  H403--H412, publisher: American Physiological Society.
\newblock \href {https://doi.org/10.1152/ajpheart.01240.2005} {\path{doi:10.1152/ajpheart.01240.2005}}.
\newline\urlprefix\url{https://journals.physiology.org/doi/full/10.1152/ajpheart.01240.2005}

\bibitem{mcgarvey_temporal_2015}
J.~R. {McGarvey}, D.~Mojsejenko, S.~M. Dorsey, A.~Nikou, J.~A. Burdick, J.~H. Gorman, B.~M. Jackson, J.~J. Pilla, R.~C. Gorman, J.~F. Wenk, \href{https://linkinghub.elsevier.com/retrieve/pii/S0003497515003963}{Temporal changes in infarct material properties: An in vivo assessment using magnetic resonance imaging and finite element simulations} 100~(2)  582--589.
\newblock \href {https://doi.org/10.1016/j.athoracsur.2015.03.015} {\path{doi:10.1016/j.athoracsur.2015.03.015}}.
\newline\urlprefix\url{https://linkinghub.elsevier.com/retrieve/pii/S0003497515003963}

\bibitem{ojha_myocardial_2025}
N.~Ojha, A.~S. Dhamoon, \href{http://www.ncbi.nlm.nih.gov/books/NBK537076/}{Myocardial infarction}, in: {StatPearls}, {StatPearls} Publishing.
\newline\urlprefix\url{http://www.ncbi.nlm.nih.gov/books/NBK537076/}

\bibitem{ennis_whole_2021}
H.~Xu, S.~A. Niederer, S.~E. Williams, D.~E. Newby, M.~C. Williams, A.~A. Young, \href{https://link.springer.com/10.1007/978-3-030-78710-3_7}{Whole heart anatomical refinement from {CCTA} using extrapolation and parcellation}, Vol. 12738, Springer International Publishing, pp. 63--70, book Title: Functional Imaging and Modeling of the Heart Series Title: Lecture Notes in Computer Science.
\newblock \href {https://doi.org/10.1007/978-3-030-78710-3_7} {\path{doi:10.1007/978-3-030-78710-3_7}}.
\newline\urlprefix\url{https://link.springer.com/10.1007/978-3-030-78710-3_7}

\bibitem{fedorov_3d_2012}
A.~Fedorov, R.~Beichel, J.~Kalpathy-Cramer, J.~Finet, J.-C. Fillion-Robin, S.~Pujol, C.~Bauer, D.~Jennings, F.~Fennessy, M.~Sonka, J.~Buatti, S.~Aylward, J.~V. Miller, S.~Pieper, R.~Kikinis, 3d slicer as an image computing platform for the quantitative imaging network 30~(9)  1323--1341.
\newblock \href {https://doi.org/10.1016/j.mri.2012.05.001} {\path{doi:10.1016/j.mri.2012.05.001}}.

\bibitem{noauthor_dzhk_nodate}
\href{https://biovat.dzhk.de/}{{DZHK} study: {BioVAT}-{HF}-{DZHK}20: {DZHK} study}.
\newline\urlprefix\url{https://biovat.dzhk.de/}

\bibitem{gavenis_safety_2023}
K.~Gavenis, \href{https://clinicaltrials.gov/study/NCT04396899}{Safety and efficacy of induced pluripotent stem cell-derived engineered human myocardium as biological ventricular assist tissue in terminal heart failure}, submitted: 2020-05-11.
\newline\urlprefix\url{https://clinicaltrials.gov/study/NCT04396899}

\bibitem{tiburcy_defined_2017}
M.~Tiburcy, J.~E. Hudson, P.~Balfanz, S.~Schlick, T.~Meyer, M.-L. Chang~Liao, E.~Levent, F.~Raad, S.~Zeidler, E.~Wingender, J.~Riegler, M.~Wang, J.~D. Gold, I.~Kehat, E.~Wettwer, U.~Ravens, P.~Dierickx, L.~W. van Laake, M.~J. Goumans, S.~Khadjeh, K.~Toischer, G.~Hasenfuss, L.~A. Couture, A.~Unger, W.~A. Linke, T.~Araki, B.~Neel, G.~Keller, L.~Gepstein, J.~C. Wu, W.-H. Zimmermann, Defined engineered human myocardium with advanced maturation for applications in heart failure modeling and repair 135~(19)  1832--1847.
\newblock \href {https://doi.org/10.1161/CIRCULATIONAHA.116.024145} {\path{doi:10.1161/CIRCULATIONAHA.116.024145}}.

\bibitem{lee_multiphysics_2016}
J.~Lee, A.~Cookson, I.~Roy, E.~Kerfoot, L.~Asner, G.~Vigueras, T.~Sochi, S.~Deparis, C.~Michler, N.~P. Smith, D.~A. Nordsletten, \href{http://epubs.siam.org/doi/10.1137/15M1014097}{Multiphysics computational modeling in \${\textbackslash}boldsymbol\{{\textbackslash}mathcal\{C\}\}{\textbackslash}mathbf\{Heart\}\$} 38~(3)  C150--C178.
\newblock \href {https://doi.org/10.1137/15M1014097} {\path{doi:10.1137/15M1014097}}.
\newline\urlprefix\url{http://epubs.siam.org/doi/10.1137/15M1014097}

\bibitem{kaiser_optimizing_2019}
N.~J. Kaiser, R.~J. Kant, A.~J. Minor, K.~L. Coulombe, \href{https://doi.org/10.1021/acsbiomaterials.8b01112}{Optimizing blended collagen-fibrin hydrogels for cardiac tissue engineering with human {iPSC}-derived cardiomyocytes} 5~(2)  887--899, publisher: American Chemical Society.
\newblock \href {https://doi.org/10.1021/acsbiomaterials.8b01112} {\path{doi:10.1021/acsbiomaterials.8b01112}}.
\newline\urlprefix\url{https://doi.org/10.1021/acsbiomaterials.8b01112}

\bibitem{shih_aging_2010}
H.~Shih, B.~Lee, R.~J. Lee, A.~J. Boyle, \href{https://www.ncbi.nlm.nih.gov/pmc/articles/PMC3031493/}{The aging heart and post-infarction left ventricular remodeling} 57~(1)  9--17.
\newblock \href {https://doi.org/10.1016/j.jacc.2010.08.623} {\path{doi:10.1016/j.jacc.2010.08.623}}.
\newline\urlprefix\url{https://www.ncbi.nlm.nih.gov/pmc/articles/PMC3031493/}

\bibitem{sanada_source_2018}
F.~Sanada, Y.~Taniyama, J.~Muratsu, R.~Otsu, H.~Shimizu, H.~Rakugi, R.~Morishita, Source of chronic inflammation in aging 5  12.
\newblock \href {https://doi.org/10.3389/fcvm.2018.00012} {\path{doi:10.3389/fcvm.2018.00012}}.

\bibitem{shinde_fibroblasts_2014}
A.~V. Shinde, N.~G. Frangogiannis, \href{https://www.sciencedirect.com/science/article/pii/S0022282813003477}{Fibroblasts in myocardial infarction: A role in inflammation and repair} 70  74--82.
\newblock \href {https://doi.org/10.1016/j.yjmcc.2013.11.015} {\path{doi:10.1016/j.yjmcc.2013.11.015}}.
\newline\urlprefix\url{https://www.sciencedirect.com/science/article/pii/S0022282813003477}

\bibitem{van_den_borne_myocardial_2010}
S.~W.~M. van~den Borne, J.~Diez, W.~M. Blankesteijn, J.~Verjans, L.~Hofstra, J.~Narula, \href{https://www.nature.com/articles/nrcardio.2009.199}{Myocardial remodeling after infarction: the role of myofibroblasts} 7~(1)  30--37, publisher: Nature Publishing Group.
\newblock \href {https://doi.org/10.1038/nrcardio.2009.199} {\path{doi:10.1038/nrcardio.2009.199}}.
\newline\urlprefix\url{https://www.nature.com/articles/nrcardio.2009.199}

\bibitem{liu_current_2019}
W.~Liu, Z.~Wang, \href{https://www.ncbi.nlm.nih.gov/pmc/articles/PMC7175293/}{Current understanding of the biomechanics of ventricular tissues in heart failure} 7~(1)  2.
\newblock \href {https://doi.org/10.3390/bioengineering7010002} {\path{doi:10.3390/bioengineering7010002}}.
\newline\urlprefix\url{https://www.ncbi.nlm.nih.gov/pmc/articles/PMC7175293/}

\bibitem{kolawole_characterizing_2025}
F.~O. Kolawole, V.~Y. Wang, B.~Freytag, M.~Loecher, T.~E. Cork, M.~P. Nash, E.~Kuhl, D.~B. Ennis, \href{https://www.nature.com/articles/s41598-025-89243-2}{Characterizing variability in passive myocardial stiffness in healthy human left ventricles using personalized {MRI} and finite element modeling} 15~(1)  5556, publisher: Nature Publishing Group.
\newblock \href {https://doi.org/10.1038/s41598-025-89243-2} {\path{doi:10.1038/s41598-025-89243-2}}.
\newline\urlprefix\url{https://www.nature.com/articles/s41598-025-89243-2}

\bibitem{pfaller_importance_2019}
M.~R. Pfaller, J.~M. Hörmann, M.~Weigl, A.~Nagler, R.~Chabiniok, C.~Bertoglio, W.~A. Wall, \href{https://doi.org/10.1007/s10237-018-1098-4}{The importance of the pericardium for cardiac biomechanics: from physiology to computational modeling} 18~(2)  503--529.
\newblock \href {https://doi.org/10.1007/s10237-018-1098-4} {\path{doi:10.1007/s10237-018-1098-4}}.
\newline\urlprefix\url{https://doi.org/10.1007/s10237-018-1098-4}

\bibitem{delicce_physiology_2025}
A.~V. Delicce, A.~N. Makaryus, \href{http://www.ncbi.nlm.nih.gov/books/NBK470295/}{Physiology, frank starling law}, in: {StatPearls}, {StatPearls} Publishing.
\newline\urlprefix\url{http://www.ncbi.nlm.nih.gov/books/NBK470295/}

\bibitem{nikou_effects_2016}
A.~Nikou, S.~M. Dorsey, J.~R. {McGarvey}, J.~H. Gorman, J.~A. Burdick, J.~J. Pilla, R.~C. Gorman, J.~F. Wenk, Effects of using the unloaded configuration in predicting the in vivo diastolic properties of the heart 19~(16)  1714--1720.
\newblock \href {https://doi.org/10.1080/10255842.2016.1183122} {\path{doi:10.1080/10255842.2016.1183122}}.

\end{thebibliography}



\section{Appendix}

\subsection{Klotz Curve}
\par To determine an appropriate scaling for material parameters based on image data and pressure data, we apply the Klotz curve with quasi-static loading. This curve is a phenomenological relationship between end-diastolic pressure and volume in the LV \cite{klotz_single-beat_2006}. In this work, we identify a linear and exponential scaling that we can optimize our scalings to both the end-diastolic point and pressure-volume curve behavior. We do this for the RV separately - since we apply separate pressure and volume data for each ventricle, and there is no known pressure-volume curve for the RV as there is for the left, we separately define linear scaling values for each ventricle and assume exponential behavior to be the same for both. Code detailing this nonlinear optimization method can be found on GitHub here \url{https://github.com/javijv4/klotzID}. 

\subsection{Data Tables}

\par See \ref{tab:cardiac_function_trans} for a full table with functional metrics for each variation. 

\begin{table}[h]
\centering
\begin{tabular}{|p{2.5cm}|c|c|c|c|c|}
\hline
\textbf{\makecell{Patch Variation}} & \textbf{\makecell{LV Stroke \\ Volume [mL] }} & \textbf{Stroke Work [mJ] } & \textbf{\makecell{LV Ejection \\ Fraction}} & \textbf{\makecell{Peak Systolic \\ Pressure [mmHg] }} \\
\hline
Healthy & 66.82 & 0.8948 & 0.3119 & 126.2 \\ \hline
Fibrotic & 57.92 & 0.6556 & 0.2787 & 112.3 \\ \hline
Basic & 58.26 & 0.6605 & 0.2675 & 112.1 \\ \hline
Aligned & 59.55 & 0.6950 & 0.2717 & 114.2 \\ \hline
Longitudinal & 57.79 & 0.6524 & 0.2645 & 111.9 \\ \hline
Stiffer & 58.16 & 0.6581 & 0.2690 & 111.9 \\ \hline
Softer & 58.39 & 0.6636 & 0.2651 & 112.2 \\ \hline
Active 20\% & 59.01 & 0.6786 & 0.2715 & 113.1 \\ \hline
Active 30\% & 59.76 & 0.6971 & 0.2755 & 114.2 \\ \hline
Active 50\% & 61.11 & 0.7316 & 0.2827 & 116.2 \\ \hline
Active 80\% & 62.65 & 0.7725 & 0.2912 & 118.8 \\ \hline
Aligned 50\% & 63.48 & 0.7972 & 0.2930 & 120.0  \\ \hline
Aligned 80\%  & 65.74 & 0.8629 & 0.3057 & 124.2  \\ \hline
Surface & 56.34 & 0.6172 & 0.2747 & 109.6 \\ \hline
Active Surface & 56.82 & 0.6289 & 0.2775 & 110.3 \\ \hline
Hyperactive \\ Surface & 57.12 & 0.6364 & 0.2792 & 110.8 \\ \hline
Soft Surface & 56.80 & 0.6443 & 0.2606 & 110.3 \\ \hline
Thinned Fibrotic & 57.43 & .6421 & 0.2602 & 111.1  \\ \hline
Thin Surface & 57.53 & 0.6447 & 0.2607 & 111.3 \\ \hline
Active Thin \\ Surface & 57.96 & 0.6552 & 0.2630 & 111.9 \\ \hline
Hyperactive Thin Surface & 58.40 & 0.6660 & 0.2654 & 112.6 \\ \hline
Thin Soft Surface & 57.51 & 0.6443 & 0.2606 & 111.2 \\
\hline
\end{tabular}
\caption{Cardiac Functional Metrics for all patch variations}
\label{tab:cardiac_function_trans}
\end{table}

\subsection{Problem details}
\par In the 3D-0D coupled simulation, we apply Lagrange Multipliers representing pressure in the right and left ventricles \cite{hirschvogel_monolithic_2017}. This pressure changes via the 0D lumped parameter representation, and it drives the volume change in the solid mechanical simulations that are then returned to the 0D lumped parameter representation. We apply Dirichlet boundary conditions on displacement at each heart valve of our biventricular mesh. We approximate the reference state of the heart using the systolic frame, which is known to change the estimated material parameters \cite{nikou_effects_2016}.

\end{document}